\documentclass[aps,prc,twocolumn,times,graphicx,tighten,
showpacs,showkeys,nofootinbib]{revtex4}
\usepackage[dvips]{graphicx}
\usepackage[dvips]{color}
\usepackage{amsmath}

\newcommand{\bea}{\begin{eqnarray}}
\newcommand{\eea}{\end{eqnarray}}
\newcommand{\be}{\begin{equation}}
\newcommand{\ee}{\end{equation}}
\newcommand{\np}{{\bf p}}
\newcommand{\hp}{\widehat{\bf p}}

\newcommand{\nk}{{\bf k}}
\newcommand{\nK}{{\bf K}}

\newcommand{\nx}{{\bf x}}
\newcommand{\nr}{{\bf r}}
\newcommand{\nR}{{\bf R}}

\newcommand{\sumint}{\sum\kern -3.5ex \int\kern 1.0ex}

\newlength\dlf  

\begin{document}

\title{
Coarse grained short-range correlations
}

\author{
I. Ruiz Simo$^a$,
R. Navarro P\'erez$^b$,
J.E. Amaro$^a$ 
and
E. Ruiz Arriola$^a$ 
}

\affiliation{$^a$Departamento de F\'{\i}sica At\'omica, Molecular y Nuclear,
and Instituto de F\'{\i}sica Te\'orica y Computacional Carlos I,
Universidad de Granada, Granada 18071, Spain}

\affiliation{$^b$Nuclear and Chemical Science Division, \\ Lawrence
Livermore National Laboratory, Livermore, CA 94551, USA}

\date{\today}


\begin{abstract}
We develop a scheme to take into account the effects of short-range
nucleon-nucleon correlations in the nucleon-pair wave function by
solving the Bethe-Goldstone equation for a coarse grained delta shell
potential in S-wave configuration.  The S-wave delta shell potential
has been adjusted to reproduce the $^1$S$_0$ phase shifts of the AV18
potential for this partial wave up to 2 GeV in the laboratory kinetic
energy.  We show that a coarse grained potential can describe the high
momentum tail of the back-to-back correlated pairs and the $G$-matrix
in momentum space.  We discuss the easiness and robustness of the
calculation in coordinate space and the future improvements and
utilities of this model.  This work suggests the possibility of using
perturbation theory for describing the short-range correlations, and
related to this, to substitute the $G$-matrix by an appropriate coarse-grained
potential.
\end{abstract}


\maketitle

\section{Introduction}

Historically the existence of a strong repulsive core was first pointed out by
Jastrow in 1950 in his analysis of proton-proton scattering (for a
historical overview covering up to 1989 we refer to
\cite{Machleidt:1989tm}). The assumption that this repulsive core
dominates short-range correlations, preventing two nucleons to
approach each other to distances closer than half a fermi, invalidates
direct use of mean field methods applied directly to the NN
interaction as inferred directly from NN scattering data. From a
Wilsonian point of view the strength of the interaction, however,
depends on the probing wavelength $\Delta r$ (see
e.g. Ref.~\cite{RuizArriola:2016vap} and references therein) which is
ultimately related to a high energy cut-off in the problem.

In this paper we want to focus on the description of high momentum
components of the nuclear wave function. Since the core distance $r_c$
and the Fermi momentum $k_F$ at saturation fulfill $2 k_F r_c \sim 1$
we will see that by judiciously tuning the corresponding length scale
$\Delta r \sim r_c $ it is possible to access high momentum pair
distributions in nuclear matter with back-to-back momentum $p \sim 2
k_F$ while keeping scattering information for the same situation in
the free space without need of a strong repulsive core. Going beyond
this maximum momentum is possible within a NN potential model
approach, but it also faces the problem that the NN interaction needs
to take into account a substantial inelasticity which would require
explicit consideration of nucleon resonance production, such as $NN
\to N \Delta$ or $NN \to \Delta \Delta$ for $p_\Delta \sim \sqrt{M
  (M_\Delta-M)} \sim 2 k_F$
in the scattering problem.

There exist many possible ways and techniques to handle and analyze
short range correlations
 (see e.g. \cite{ring2004nuclear} and references therein). 
For our purposes we will follow the venerable
Brueckner-Goldstone \cite{Brueckner:1954zz, Brueckner:1958zz,
  Brueckner:1955zze} theory of nuclear matter. We believe this
framework is a good starting point which provides a satisfactory method for finding 
the bulk properties of nuclear matter, in particular its saturation
energy and equilibrium density.

The formal derivation and the mathematical framework to carry out
calculations within this theory were also provided by Goldstone and
Bethe \cite{Goldstone:1957zz, Bethe:1956zz} in terms of an
integro-differential equation in configuration space.  Excellent
reviews on this topic can be found on Refs.  \cite{Bethe:1971xm,
  Day:1967zza, Brown:1971zza}.

One of the main ingredients in solving the G-matrix in the
Brueckner-Bethe-Goldstone formalism is the two-nucleon potential
$V_{ij}$. Most early and modern so-called realistic nucleon-nucleon
(NN) potentials \cite{Wiringa:1994wb, Reid:1968sq, Machleidt:2000ge,
  Hamada:1962nq, Nagels:1977ze, Nagels:1978sc,Lacombe:1980dr} contain
parameters that are usually fitted to available scattering
information  up to a certain
energy (usually about pion production threshold), and they also
provide the right static properties of the unique bound state of the
two-nucleon system, namely the deuteron.

The AV18 potential~\cite{Wiringa:1994wb} is a popular and versatile
choice which has attracted much attention from nuclear structure
theorists 
since it described np and pp scattering with
   $\chi^2/\rm dof \sim  1.1$ for LAB energies up to 350MeV at the time of the Nijmegen analysis
   (the quality has worsened with the new Granada database to  $\chi^2/\rm dof \sim 1.46$
    without refitting up to 300 MeV due to the 40\% more new data~\cite{Pia15}). 
Besides being local, it presents both a repulsive core and turns out
to provide {\it a fortiori} a qualitative high energy description of
scattering data when the inelasticity effects are neglected. Our
approach in this work will consist in taking a coarse grained delta
shell potential whose strength parameters are fitted to obtain the
same phase shifts as the AV18 potential \cite{Wiringa:1994wb} up to a
laboratory kinetic energy of 2 GeV.  The coarse graining is based on
the idea that if one wants to determine some NN scattering observables
in a limited energy range, then the interaction potential only needs
to be known in a limited number of points. The average separation of
these points is related to the maximum resolution power which can be
achieved within this upper limited energy range implying a shortest de
Broglie wavelength. The findings of this work are based on the early
an insightful work by Avil\'es~\cite{Aviles:1973ee} which was
rediscovered within a Wilsonian renormalization
perspective \cite{Entem:2007jg} and fully exploited in NN scattering
analysis \cite{NavarroPerez:2011fm, NavarroPerez:2012qf,
NavarroPerez:2012qr, Perez:2013mwa, Perez:2013jpa,Perez:2014waa} 
to which we refer
for further details.  We emphasize that this simplification takes
place in configuration space and we will exploit this feature
explicitly in our analysis.

Our aim is to study the properties of the coarse-grained
(GR) potential in the nuclear medium.  In the past we studied and
fitted the GR to NN scattering data. Here we go beyond the bare NN
interaction and study the influence of coarse-grain on the
short-range correlations between nucleon pairs inside the nucleus, by
analyzing the high-momentum components of the relative wave function,
comparing with the potential AV18, which produce similar
phase-shifts.

Since the appearance of the Bethe-Goldstone (BG) equation, several
methods have been developed to solve it (for a critical review on some
of them we refer the reader to Ref.  \cite{Dahll:1969hmo}).  Most of
the early treatments analyzed the problem in configuration space until
Haftel and Tabakin introduced a momentum space solution via a direct
matrix inversion method after some smoothing of NN interaction was
implemented~\cite{Haftel:1970zz}. It should be noticed that a
repulsive core in configuration space generates long high momentum
tails which inevitably lead to large matrices 
(typical dimensions are of the order of  $\sim 50$) \cite{Lan92}.  While it is
possible to carry out such a coarse grained analysis we will proceed
here directly by using the original integro-differential version of the
BG equation.

Besides, some inherent difficulties in the BG solution have been
overcome by resorting to approximate solutions. Just to enumerate a
few of them, we can mention: the treatment of the
center-of-mass (CM) motion of the two-nucleon system; the handling of
the Pauli blocking operator \cite{Cheon:1988hn} or the necessity (or
not) of a partial wave decomposition \cite{White:2013xxa}.  The first
two of the above difficulties are easily overcome by choosing the
kinematic configuration where the CM momentum of the nucleon pair is
zero. This situation corresponds to a back-to-back configuration for
the correlated nucleon pair, and recently it has drawn attention from
the theoretical nuclear physics community \cite{Feldmeier:2011qy,
  Vanhalst:2012ur, Alvioli:2013qyz, Niewczas:2015iea, Mosel:2016uge,
  VanCuyck:2016fab, Weinstein:2016inx} as well as from the
experimental electron \cite{Marchand:1987hd, Onderwater:1998zz,
  Blomqvist:1998gq, Egiyan:2005hs, Shneor:2007tu} and neutrino
scattering physics communities \cite{Acciarri:2014gev,
  Cavanna:2015sla}.

In this work we try to keep the maximum simplicity as possible in the
approach to this problem in order to properly understand the effects
of all the ingredients involved and how are they mutually
intertwined. Therefore we try to be pedagogical and refrain from going
beyond the S-wave configuration in a partial wave expansion. We also
restrict our calculation to the back-to-back configuration for a
nucleon pair at rest ($\mathbf{P_{CM}=0}$) in order to avoid all the
problems related to the CM motion, specifically the angular averaging
of the Pauli blocking operator.

The structure of the paper is as follows.  In section II we review the
Bethe-Goldstone equation in coordinate representation for the
correlated wave function. As originally recognized by Bethe and
Goldstone this is particularly suited when there is a strong repulsive
piece of the interaction. We also deal with the correlated wave
function in momentum representation before entering, in section III,
into the discussion of the results on the high momentum tail of the
momentum distribution, the $G$-matrix, and how the repulsive and
attractive components are reshuffled depending on the resolution
wavelength. Finally we summarize our findings and outline our future
plans in section \ref{sec4}.

\section{Formalism}

The Bethe-Goldstone equation can be regarded as the in-medium
scattering equation. This is an integral equation which, besides the
effect of the inter-particle potential, also incorporates the effect
of the surrounding medium by preventing the interacting nucleon pair
from being scattered into the already occupied levels below the Fermi
momentum $k_F$. Here we review it so that our conventions and notation
as well as our coarse grained method of solution and analysis can be
more easily introduced in configuration space. While the original BG
discussion has become textbook material~\cite{FetWal03,Wal95} only
some simple cases were discussed in a rather sketchy fashion.  We hope
that our presentation will be useful both for newcomers as well as
researchers familiarized with the more popular momentum space approach.

\subsection{Correlated wave function} \label{subsec2a}

The  Bethe-Goldstone equation for the wave function of an interacting
nucleon pair in the independent-pair approximation \cite{Wal95} can be
written in the form
\begin{eqnarray}
&& \left|\Psi\right\rangle = \left|\nk_1 \nk_2\right\rangle +
 \int d^3\nk^\prime_1 d^3\nk^\prime_2\;
 \left|\nk^\prime_1 \nk^\prime_2\right\rangle
\frac{\left\langle \nk^\prime_1 \nk^\prime_2\right|
V\left|\Psi\right\rangle}
{E-(E_{\nk^\prime_1}+E_{\nk^\prime_2})}\nonumber\\
&& \times\; \theta(|\nk^\prime_1|>k_F)\;
 \theta(|\nk^\prime_2|>k_F)
\label{BG_wafunc}
\end{eqnarray}
where $\theta(x)$ is the Heaviside or unit step function, 
and $\left|\Psi\right\rangle$ is the correlated
 wave function. This is given in terms of the unperturbed 
plane wave solution $\left|\nk_1 \nk_2\right\rangle$ (with
$\left|\nk_i\right|<k_F$) plus the high momenta 
components $\left|\nk^\prime_1 \nk^\prime_2\right\rangle$
(with $\left|\nk^\prime_i\right| > k_F$), which are 
weighted by the transition matrix element of the potential 
and the energy denominator that damps the largest
energy differences with respect to the exact eigenvalue
$E$. The unperturbed energies $E_{\nk^\prime_i}$
correspond to the eigenvalues of the unperturbed
Hamiltonian $H_0=T_1+T_2$ and are given by
\begin{equation}\label{kinetic_energy}
E_{\nk^\prime_i}=\frac{\nk^{\prime 2}_i}{2 M_N}
\end{equation}
where $M_N$ is the nucleon mass.

Note that in eq. (\ref{BG_wafunc}) we have particularized 
the solution of the B-G equation for the case of nuclear 
matter or a Fermi gas, where the plane wave solutions
with definite momenta are known to be the eigenfunctions
of the uncorrelated system.

If we left-multiply eq. (\ref{BG_wafunc}) by the
bra $\left\langle \nx_1 \nx_2\right|$, we get the
correlated wave function in coordinate representation as
\begin{eqnarray}
&&\Psi(\nx_1,\nx_2)=\Phi_{\nk_1 \nk_2}(\nx_1,\nx_2)
+ \int d^3\nk^\prime_1 d^3\nk^\prime_2\nonumber\\
&&\times\,\Phi_{\nk^\prime_1 \nk^\prime_2}(\nx_1,\nx_2)\;
\theta(|\nk^\prime_1|>k_F)\;
 \theta(|\nk^\prime_2|>k_F)
\nonumber\\
&\times&\frac{\int d^3\nx^\prime_1 d^3\nx^\prime_2 \;
\Phi^*_{\nk^\prime_1 \nk^\prime_2}
(\nx^\prime_1,\nx^\prime_2)\; V(\nx^\prime_1,\nx^\prime_2)\;
\Psi(\nx^\prime_1,\nx^\prime_2)}
{E-(E_{\nk^\prime_1}+E_{\nk^\prime_2})}
\label{BG_wafunc_coor}
\end{eqnarray}
where it is easier to notice the self-consistent integral
character of the B-G equation.

Now we use the trick of changing all coordinate and momenta
from particles' variables to CM and relative ones
\begin{eqnarray*}
\nR_{\rm CM} =\frac12(\nx_1+\nx_2)&;& \qquad 
\nK_{\rm CM}=\nk_1+\nk_2\\
\nx = \nx_1-\nx_2&;& \qquad \nk=\frac12(\nk_1-\nk_2)
\end{eqnarray*}
Furthermore, if the  potential only depends
on the relative coordinate $\nx$ and not on the CM one, 
further simplifications are possible. This
kind of potentials preserve the CM motion and the 
correlated total wave function is separable into a product
of a plane wave describing the free motion of the CM 
system and a correlated relative wave function $\psi$ that
fulfills a BG-like equation \cite{Wal95}
\begin{eqnarray}
\Psi(\nx_1,\nx_2)
&=&
\frac{e^{i\nK_{\rm CM}\cdot \nR_{\rm CM}}}{(2\pi)^{\frac32}}
\frac{\psi_{\nK_{\rm CM},\nk}(\nx)}{(2\pi)^{\frac32}}
\label{total_wafunc}\\
\Phi_{\nk_1 \nk_2}(\nx_1,\nx_2)
&=&
\frac{e^{i\nK_{\rm CM}\cdot \nR_{\rm CM}}
}{(2\pi)^{\frac32}}\,
\frac{e^{i\nk\cdot \nx}
}{(2\pi)^{\frac32}}\,
\label{unperturbed_wafunc}
\end{eqnarray}
and analogously for primed variables.

If the above expressions (\ref{total_wafunc}) and 
(\ref{unperturbed_wafunc}) are substituted into eq.
(\ref{BG_wafunc_coor}) then the CM wave functions
factorize and cancel on both sides and 
we are left with the BG-like equation
for the correlated relative wave function $\psi$,
\begin{eqnarray}
&&\psi_{\nK_{\rm CM},\nk}(\nx)=
e^{i\nk\cdot \nx}+\int \frac{d^3\nk^\prime}{(2\pi)^3}\,
e^{i\nk^\prime\cdot \nx}\; \frac{1}
 {\frac{\nk^2}{2\mu}-\frac{\nk^{\prime 2}}{2\mu}}\nonumber\\
&&\theta\left(\left|\frac{\nK_{\rm CM}}{2}-
\nk^\prime\right|-k_F\right)\theta\left(\left|
\frac{\nK_{\rm CM}}{2}+\nk^\prime\right|-k_F\right) \nonumber\\ 
&& \int d^3\nx^\prime\, e^{-i\nk^\prime\cdot \nx^\prime}\,
  V(\nx^\prime)\,\psi_{\nK_{\rm CM},\nk}(\nx^\prime)
  \label{relative_BG_wafunc}
\end{eqnarray}
where $\mu=\frac{M_N}{2}$ stands for the reduced mass
of the two-nucleon system; the energy denominator now only
contains the difference between the initial and final
relative kinetic energies (the initial and final CM energies cancel
in the denominator). The two step functions impose the 
conditions $|\nk^\prime_1|,|\nk^\prime_2|>k_F$.

One of the main complications in solving 
eq. (\ref{relative_BG_wafunc}) involves the integration 
over the angles of $\nk^\prime$ due to the presence of 
the step functions, which explicitly depend on the angle
between $\nK_{\rm CM}$ and $\nk^\prime$. For that
reason
an angular average of the
projection operator 
is usually performed 
\cite{Brueckner:1958zz} (see Ref. 
\cite{Wer59} for a full treatment of this problem in the
general case). Another possibility is simply 
considering the situation where
 $\nK_{\rm CM}=\mathbf{0}$, which corresponds to a 
 back-to-back configuration for the two-nucleon system
(where the effect of NN correlations should be maximized).
 Note that in the latter case the two step functions get reduced
 to a single one, simply implying that $|\nk^\prime|>k_F$.

\subsection{Integro-differential Bethe-Goldstone equation} \label{subsec2axx}

Up to this point the discussion has been quite general on
the correlated wave function problem. Next discussion can 
be also followed from Refs \cite{Wal95, Bet57}, where the
integral B-G equation for the relative wave function 
(\ref{relative_BG_wafunc}) is transformed in an 
integro-differential equation by applying the operator
$(\nabla^2+\nk^2)$ on both sides of eq. 
(\ref{relative_BG_wafunc}) and taking the especial case
when $\nK_{\rm CM}=\mathbf{0}$. Indeed, we have then
\begin{eqnarray}
&&(\nabla^2+\nk^2)\psi_{\nk}(\nx)=
\int\frac{d^3\nk^\prime}{(2\pi)^3}\;
e^{i\nk^\prime\cdot\nx}\; \theta\left(\left|
\nk^\prime\right|-k_F\right)
\nonumber\\
&\times&\int d^3\nx^\prime\; 
e^{-i\nk^\prime\cdot\nx^\prime}\; 2\mu\; V(\nx^\prime)\;
\psi_{\nk}(\nx^\prime)
\\
&=& 2\mu\, V(\nx)\, \psi_{\nk}(\nx)-
\int\frac{d^3\nk^\prime}{(2\pi)^3}\;
\theta\left(k_F-\left|\nk^\prime\right|\right)\,
e^{i\nk^\prime\cdot\nx}
\nonumber\\
&\times&\int d^3\nx^\prime\; 
e^{-i\nk^\prime\cdot\nx^\prime}\; 2\mu\; V(\nx^\prime)\;
\psi_{\nk}(\nx^\prime)\label{integro_diff_eq}
\end{eqnarray}
where in the last steps we have used the property
of the Heaviside function $\theta(x)=1-\theta(-x)$
and the orthonormality condition of the plane waves
\begin{equation*}
\int d^3\nk^\prime\; e^{i\nk^\prime\cdot(\nx-\nx^\prime)}
=(2\pi)^3 \, \delta^3(\nx-\nx^\prime)
\end{equation*}

Now we expand the wave function $\psi_{\nk}(\nx)$ in partial waves.
For simplicity in this work we assume that the nucleon pair is coupled
to total spin $S=0$. Then the NN potential do not mix different
partial waves. The function
$V_l(s)$ is the $l$-multipole of the potential in the $^1L_L$ 
 channel, and each reduced radial wave function, $u_{k,l}(r)$,
verifies the decoupled
integral-differential equation:
\begin{eqnarray}
&&\left\lbrace \frac{d^2}{dr^2}-\left(\frac{l(l+1)}{r^2}-
\left(\nk^2-2\mu\, V_l(r)  \right)  
\right)\right\rbrace u_{k,l}(r)=\nonumber\\
& -&\frac{4\mu\,r}{\pi}\int^{k_F}_{0} dk^\prime\,
k^{\prime 2}\, j_l(k^\prime r) \int^{\infty}_{0}
dr^\prime\, r^\prime\,  j_l(k^\prime r^\prime)
V_l(r^\prime)u_{k,l}(r^\prime)\nonumber\\
\label{integro_diff_radial_wf}
\end{eqnarray}
where $r=\left|\nx\right|$ is the 
relative distance, $k=\left|\nk\right|$ and 
$ j_l(\rho)$ is a spherical Bessel function.

From now on we particularize eq. (\ref{integro_diff_radial_wf})
for the S-wave case ($l=0$).  We have the BG equation which is solved in Sect. IIE,
\begin{eqnarray}
&&\left\lbrace \frac{d^2}{dr^2}+
\left(\nk^2-2\mu\, V(r)  \right)  \right\rbrace 
u_{k}(r)=\nonumber\\
& -&\frac{4\mu}{\pi}\int^{\infty}_{0}
dr^\prime\, 
\chi(r,r^\prime)
V(r^\prime)\; u_{k}(r^\prime)\;
\label{integro_diff_radial_wf_swave}
\end{eqnarray}
where the kernel $\chi(r,r^\prime)$ is given by
\cite{Wal95}
\begin{eqnarray}
&&\chi(r,r^\prime)=\int^{k_F}_{0} dk^\prime\,
\sin(k^\prime r)\, \sin(k^\prime r^\prime)\nonumber\\
&&= \frac12\left[\frac{\sin\left(k_F(r-r^\prime)\right)}{r-r^\prime}-
\frac{\sin\left(k_F(r+r^\prime)\right)}{r+r^\prime} \right].
\label{kernel_chi}
\end{eqnarray}

\subsection{Coarse graining vs fine graining} \label{subsec2axxx}

The novelty of the present work consists in solving eq.
(\ref{integro_diff_radial_wf_swave}) for equally spaced delta shell
potential. As such, this method can be regarded as a simple quadrature
method for a {\it given} potential $V(r)$, namely making the
replacement
\begin{equation}
V(r) \to \sum^{N}_{i=1} \Delta r V(r_i) \;\delta(r-r_i) 
\label{deltashell_pot_fine}
\end{equation}
for an equidistant grid $r_n = n \Delta r$, which stops when $r_N \sim
a$ with $a$ the range of the interaction. Of course, for $N \to
\infty$ and $\Delta r \to 0$, we expect a better and more accurate
solution. Here $\Delta r$ plays the role of an integration step for
this particular quadrature method.  We will show below that this {\it
  fine graining} requires in practice a large number of mesh points
for a potential $V(r)$ such as the AV18 which has been determined from
a fit to NN scattering data up to a maximum energy.

In contrast, the {\it coarse graining} approach already presented in
Refs. \cite{Aviles:1973ee, NavarroPerez:2011fm, NavarroPerez:2012qf,
  NavarroPerez:2012qr, Perez:2013mwa, Perez:2013jpa} corresponds to
take $\Delta r$ not as an auxiliary integration step but as a physical
parameter. Namely, we take it as the shortest de Broglie wavelength
resolution $\Delta r \sim \lambda_{\rm min}=1/p_{\rm CM}^{\rm max}$
with $p_{\rm CM}^{\rm max}$ the maximum CM momentum and use the values
of $V_{\Delta r}(r_i)$ as fitting parameters themselves, thus we make
the replacement
\begin{equation}
V(r) \to \sum^{N}_{i=1} \Delta r V_{\Delta r}(r_i) \;\delta(r-r_i)
 \equiv 
\sum^{N}_{i=1} \frac{\lambda_i}{2\mu}\;\delta(r-r_i)
\label{deltashell_pot}
\end{equation}
Obviously the number $N$ and values of these fitting parameters {\it
  depend} on the resolution wavelength, $\Delta r$ and hence on the
maximal fitting energy as well as the desired accuracy in the
phase-shifts. For a potential with range $a$ we expect $N \sim a/
\Delta r = a p_{\rm CM}^{\rm max}$ points. Thus, if we take a
potential $V(r)$ with phase shifts $\delta(p)$, its coarse grained
representation $V_{\Delta r} (r)$, corresponds to find $V_{\Delta r}
(r_i)$ such that $\delta_{\Delta r} (p)= \delta (p) \pm \Delta \delta
(p)$ for $p \le 1/\Delta r$ and with $\Delta \delta (p)$ the tolerated
discrepancy. In practice we use the standard $\chi^2$ as a figure of merit 
\begin{eqnarray}
\chi^2 ( \lambda_1, \dots, \lambda_N) 
= \sum_{n=1}^{N_p} \left[\frac{\delta_{\Delta r} (p_n) - \delta (p_n)
}{\Delta \delta (p_n)} \right]^2   
\end{eqnarray}
and determine the coarse grained parameters $\lambda_i$ by
minimization.  An educated guess is to take $\Delta \delta (p)$ as the
expected systematic discrepancies \cite{Perez:2014waa} and $p_n$ as the
values corresponding to the tabulated LAB energies.

Of course, we expect that for $\Delta r \to 0 $, $V_{\Delta r}
(r_i) \to V(r_i)$~\footnote{Note that this is not the same as fixing
  $p_{\rm CM}^{\rm max}$ and increasing the wavelength resolution by
  taking $\Delta r p_{\rm CM}^{\rm max} \ll 1$. In this case, the
  final potential, while continuous, does not reproduce the original
  one; oscillations of period $\sim 2\pi/p_{\rm CM}^{\rm max}$ develop
  exhibiting the physical resolution.}. These considerations hold
equally well {\it regardless} on the value of the Fermi momentum
$k_F$. In order to illustrate the procedure we will discuss first the
vacuum case, $k_F=0$, corresponding to the scattering problem 
before coming to the BG solution.

\subsection{Scattering Solution for $\delta$-shell potentials} 
\label{subsec2axxxx}

The scattering problem in the $S$-wave corresponds to
solving the equation \cite{Entem:2007jg,NavarroPerez:2011fm}
\begin{eqnarray}
&&\left\lbrace \frac{d^2}{dr^2}+
\left(\nk^2-\sum^{N}_{i=1} \lambda_i\;\delta(r-r_i)  
\right)  \right\rbrace 
u_{k}(r)=0\label{scattering_ODE}
\end{eqnarray}
with the boundary conditions at the origin and at infinity 
\begin{eqnarray}
u_k(0)=0 \, , \qquad  u_k (r) \to C_k \sin (k r + \delta(k))
\end{eqnarray}
The solution outside anyone
of the concentration radii ($r_i$) can be written as 
\begin{eqnarray}
&&u_{k}(r)=A_i\sin(kr+\delta_i)\label{scattering_sol}   \\
&&{\rm for} \qquad r_i < r < r_{i+1} \quad (i=0, 1,\cdots N)\nonumber
\end{eqnarray}
where $\delta_i$ is the accumulated phase shift due to the delta
shells potential up to $r_i$ and $A_i$ are amplitudes which are fixed
by an arbitrary normalization condition (for instance $A_N=1)$. Taking
$r_0=0$ and $\delta_0=0$ the phase shift is given by the total
accumulated one
\begin{eqnarray}
\delta (k) = \delta_N 
\end{eqnarray}
Now it is necessary to match the different branches of the reduced
wave function on each interval around the $i$-th delta shell, and this
is done by imposing continuity of the total wave function at the
$i$-th concentration radius and discontinuity of the first derivative
of the wave function due to the existence of an extremely singular
potential at $r=r_i$. These two conditions reduce to
  \begin{eqnarray}
 && u_k(r^+_{i}) = u_k(r^-_{i})\label{continuity_cond}\\
 && u^\prime_k(r^+_i) - u^\prime_k(r^-_i) = \lambda_i u_k(r_i) \quad
 (i=1, 2 \cdots N) \label{disc_condition_first_der}
  \end{eqnarray}
  where $r^{\pm}_i=\lim_{\epsilon\rightarrow 0^{{}^{+}}} (r_i \pm \epsilon)$
  and the prime over $u_k$ here denotes the first derivative.
The condition expressed in eq. (\ref{disc_condition_first_der})
can be straightforwardly obtained by using the solution  
around any of the concentration radii $r_i$, Eq.~(\ref{scattering_sol}),
yielding the simple recurrence relation at the point $r_{i+1}$
\begin{eqnarray}\label{recurrence_phaseshifts}
k \cot ( k r_{i+1} + \delta_{i+1} )- k \cot ( k r_{i+1} + \delta_i ) 
= \lambda_{i+1} 
\end{eqnarray}
whence the total accumulated phase-shift may be computed.  The
recurrence relation for the amplitudes of the homogeneous solution is
then
\begin{equation}\label{recurrence_amplitudes}
A_{i+1}=A_i\;\frac{\sin(kr_{i+1} + \delta_i)}{\sin(kr_{i+1}+{\delta}_{i+1})}
\end{equation}


\subsection{BG Solution for $\delta$-shell potentials} \label{subsec2axxxxx}

We come here to the core of our construction.  The Bethe-Goldstone
equation (\ref{integro_diff_radial_wf_swave}) 
is a linear integral-differential equation. For the 
 $^1S_0$ delta shell potential
it reads
\begin{equation}
\left\lbrace \frac{d^2}{dr^2}+
\left(k^2-\sum^{N}_{i=1} \lambda_i\;\delta(r-r_i)  
\right)  \right\rbrace 
u_{k}(r)= F_k(r)
\label{diff_eq_delta_ri}
\end{equation}
where we have defined the source function, $F_k(r)$,
which  depends on the
values of the wave function $u_{k}(r_i)$ at the 
points $r_i$ where the delta shells are localized
\begin{equation}
\label{source}
F_k(r)\equiv -\frac{2}{\pi}\sum^{N}_{i=1} \lambda_i \; 
u_{k}(r_i)\; \chi(r,r_i).
\end{equation}
If we assume that the values $u_k(r_i)$ are known, the
above equation (\ref{diff_eq_delta_ri})
 corresponds to a second-order ordinary
differential equation (ODE) with a source term.
Therefore 
$u_k(r)$  can be written as the sum 
of a solution of the homogeneous ODE plus a particular
solution of the whole ODE,
\begin{equation}
u_{k}(r)=u_{h}(r)+u_{p}(r)\label{total_solution}
\end{equation}
where $u_h(r)$ is a solution of the homogeneous equation
\begin{equation}
\left\lbrace \frac{d^2}{dr^2}+
\left(k^2-\sum^{N}_{i=1} \lambda_i\;\delta(r-r_i)  
\right)  \right\rbrace 
u_{h}(r)=0.
\label{homogeneous_ODE}
\end{equation}
The condition of regularity at origin imposes $u_{k}(0)=0$.
The solution of
eq. (\ref{homogeneous_ODE}) outside anyone of the concentration radii
($r_i$) can be written as in Eq.~(\ref{scattering_sol}), i.e. 
\begin{eqnarray}
&&u_{h}(r)=A_i\sin(kr+\bar{\delta}_i)\label{homogeneous_sol}   \\
&&{\rm for} \qquad r_i < r < r_{i+1} \quad (i=0, 1,\cdots N). 
\nonumber
\end{eqnarray}
A particular solution $u_p(r)$ of eq.  (\ref{diff_eq_delta_ri}) 
is given by
 \begin{equation}\label{particular_solution}
 u_p(r)=\frac{1}{k}\int^{r}_{0} dr^\prime\, F_{k}(r^\prime)\,
 \sin\left(k(r-r^\prime)\right)
 \end{equation}
 which can be proven by direct differentiation  \cite{Wal95}. 
Therefore, we can write the complete solution of the
Bethe-Goldstone equation (\ref{diff_eq_delta_ri}) as the piecewise function
\begin{equation}
\label{whole_solution}
u_k(r)=A_i \sin(kr + \bar{\delta}_i)+u_p(r)\qquad {\rm if} \quad r_i <r<r_{i+1}
\end{equation}
Notice that $u_p(r)$ vanishes at origin by
construction, and that it is a
continuous function because it is the integral of the product of two
continuous functions. The regularity
condition of $u_k(r)$ at $r=0$ implies that the phase shift at origin
${\bar{\delta}}_0=0$. 

By imposing the two conditions expressed in eqs. 
(\ref{continuity_cond}) and (\ref{disc_condition_first_der})
at every point where the delta shells are localized,
we can recursively relate the amplitude and phase shift of 
the wave function on the right of each delta shell with those
of the wave function at the left of that delta shell.
For the amplitudes the recurrence relation is again
\begin{equation}
A_{i+1}=A_{i}\;\frac{\sin(kr_{i+1}+\bar{\delta}_i)}
{\sin(kr_{i+1}+\bar{\delta}_{i+1})}.
\end{equation}
The corresponding recurrence relation for the phase shifts
is given by
\begin{eqnarray}
&&k\cot(kr_{i+1}+\bar{\delta}_{i+1})-
k\cot(kr_{i+1}+\bar{\delta}_{i})=\nonumber\\
&&\lambda_{i+1}
\left(1+\frac{u_p(r_{i+1})}{A_i\sin(kr_{i+1}+\bar{\delta}_i)}\right).
\end{eqnarray}
It reduces to expression (\ref{recurrence_phaseshifts})
when there is no medium and
therefore $u_p(r)=0$, as it is the case for the free scattering problem.

Solving for $\bar{\delta}_{i+1}$ we get
\begin{eqnarray}
&&\bar{\delta}_{i+1}
= -kr_{i+1}
\nonumber\\
&&+
{\rm atan}
\Bigl\{
A_i k \sin(kr_{i+1}+\bar{\delta}_i)
\bigl[ 
A_i k\cos(kr_{i+1}+\bar{\delta}_i)
\nonumber\\
&&
+
A_i \lambda_{i+1}\sin(kr_{i+1}+\bar{\delta}_i)
+\lambda_{i+1}\,u_p(r_{i+1})
\bigr]^{-1}
\Bigr\}
\end{eqnarray}

Note that for the determination of the amplitudes $A_i$
and phase shifts $\bar\delta_i$ of the BG solution, 
it is completely necessary
to know the particular solution $u_p(r_i)$ at the points 
$r_i$ where the delta shells are located. 

Therefore we cannot get rid of the self-consistency
problem inherent to any solution of the B-G equation:
to determine the whole solution we need some
constants which have to be determined by means of recurrence
relations provided the particular solution is already known;
and to determine the particular solution, eq. 
(\ref{particular_solution}), we  need to know
 the whole  wave function $u_k(r_i)$, at the points $r_i$,
in order  to compute the source function 
$F_k(r)$ in eq. (\ref{source}).

Here we solve this problem by carrying out an iterative procedure. For
a given $k$ we take as starting points $u^{(0)}_k(r_i)=\sin(kr_i)$ to
compute the source function and the particular solution.  We then
calculate the amplitudes $A_i$ and phase shifts $\bar{\delta}_i$ to
obtain the first-iteration $u^{(1)}_k(r_i)$ of the BG solution.  With
this complete solution, we calculate again the source function and
iterate the procedure a number $N$ of times up to achieving
convergence for the $N$-th iteration  $u^{(N)}_k(r_i)$.

Because the BG equation is linear, its solution is determined up to a
normalization constant.  In our approach, in each
iteration the value of $A_0$
 is the  global normalization constant that is
fixed by imposing the long-distance condition
\begin{equation}
u_k(r) \longrightarrow \sin(kr) , \kern 1cm r \rightarrow \infty.
\end{equation}
That is, the correlated wave function of the nucleon pair, $u_k(r)$, 
approach the free relative wave function, $\sin(kr)$, at large  distances.

A convergence criterion for the iterations can be derived
from the 
solution, Eq (\ref{whole_solution}), for $r > r_N$.
In fact by expanding the particular solution (\ref{particular_solution})
in terms of sine and cosine functions, we get
\begin{eqnarray}
&&u_k(r)=A_N \sin\left(kr+\bar\delta_{N}\right)+u_p(r)
\nonumber\\
&&=\sin(kr)\left(A_N \cos\bar\delta_{N}+\frac{1}{k} 
\int^{r}_{0} dr^\prime F_k(r^\prime)
\cos(kr^\prime)  \right)\nonumber\\
&&+\cos(kr)\left(A_N\sin\bar\delta_{N} -\frac{1}{k} 
\int^{r}_{0} dr^\prime F_k(r^\prime)
\sin(kr^\prime) \right)\nonumber\\
&&\longrightarrow \quad \sin(kr) \qquad 
{\rm when}\quad r\rightarrow \infty
\label{convergence}
\end{eqnarray}
The above expansion allows us to identify the coefficients
of $\sin(kr)$  and $\cos(kr)$ for $r\rightarrow \infty$, 
thus providing the convergence
conditions
\begin{eqnarray}
&&\frac{1}{k} 
\int^{\infty}_{0} dr^\prime F_k(r^\prime)
\cos(kr^\prime)  = 1- A_N \cos\bar\delta_{N}
\label{convergence_coseno}\\
&&\frac{1}{k} \int^{\infty}_{0} dr^\prime F_k(r^\prime)
\sin(kr^\prime)=A_N \sin\bar\delta_{N}
\label{convergence_seno}
\end{eqnarray}

In each iteration we check if the solution of the B-G equation
simultaneously verify eqs.  (\ref{convergence_coseno}) and
(\ref{convergence_seno}) within a given accuracy. This is our
convergence criterion, which of course depends on "where" we have
defined the large distances behavior. In our case we have set it to
$r_{\rm max}=50$ to 100 fm and we have found that our results do not
depend on this choice. Typically we need less than six iterations to
obtain convergence for all the NN potentials considered in this work.

\subsection{High-momentum components of the pair wave function}
\label{subsec2b}

Once the Bethe-Goldstone has been solved in coordinate space it is
straightforward to 
 compute the high momentum components of the correlated pair.
We start by writing the BG equation for the relative wave function 
of a back-to-back correlated pair with total spin $S=0$, in the form
\begin{equation}
|\psi_{\nk}\rangle= 
|\nk\rangle + \int d^3p \theta(p-k_F)
\frac{1}{\frac{k^2}{2\mu}-\frac{p^2}{2\mu}}
|\np\rangle
\langle \np |V|\psi_{\nk}\rangle
\end{equation}
where $k<k_F$.  Thus in presence of the medium the interacting
pair acquires high momentum components given by the integrand in the above
equation, namely, for $p> k_F$,
\begin{equation} \label{high-momentum}
\langle\np|\psi_{\nk}\rangle= 
\frac{1}{\frac{k^2}{2\mu}-\frac{p^2}{2\mu}}
\langle \np |V|\psi_{\nk}\rangle.
\end{equation}
in this equation a general multipole expansion can be done. Here
we study the particular case of the $S$-wave contribution. 
The $l=0$ partial wave of
the BG wave function is written as
\begin{equation}
\psi_{l=0}(\nr) = \frac{1}{(2\pi)^{3/2}}\frac{u_k(r)}{kr}
\end{equation}
where $u_k(r)$ is the BG solution in $S$-wave obtained in the last
section, Eq. (\ref{diff_eq_delta_ri}).  Note that here we are using
the same normalization as the plane wave $|\nk\rangle$ which for $l=0$
is
\begin{equation}
\left[\frac{1}{(2\pi)^{3/2}}
{\rm e }^{i\nk\cdot\nr}
\right]_{l=0}
=\frac{1}{(2\pi)^{3/2}}\frac{\sin(kr)}{kr}.
\end{equation}
Therefore the matrix element of the NN potential between the
correlated  $S$-partial wave and the high-momentum state, $p>k_F$,  is
\begin{equation}
\langle \np |V|\psi_{l=0}\rangle = 
\int d^3 r 
\frac{{\rm e }^{-i\np\cdot\nr}}{(2\pi)^{3/2}}
\sum_i \frac{\lambda_i}{2\mu}\delta(r-r_i)
 \frac{1}{(2\pi)^{3/2}}\frac{u_k(r)}{kr}
\end{equation}
where we have introduced the delta-shell potential in $S$-wave as a
sum of delta functions.  The angular integral selects the $l=0$
component of the plane wave through the general relation, for any
radial function $f(r)$,
\begin{equation}
\int d^3r
{\rm e }^{-i\np\cdot\nr}f (r) = 
4\pi \int dr r^2 j_0(pr) f (r)
\end{equation}
where $j_0(pr)= \sin(pr)/pr$  is a spherical Bessel function.
Integrating over the radial coordinate using the Dirac delta functions
we obtain 
\begin{equation}\label{matrix-element}
\langle \np |V|\psi_{l=0}\rangle = 
\frac{4\pi}{(2\pi)^{3}}
\sum_i \frac{\lambda_i}{2\mu}
\frac{\sin(pr_i)}{p}
\frac{u_k(r_i)}{k}
\end{equation}
Using this result in Eq. (\ref{high-momentum}) we obtain the 
high-momentum components of the $S$-wave correlated back-to-back pair 
 in the analytical form
\begin{equation} \label{momentum-wave}
\langle \np|\psi_{l=0}\rangle = 
\frac{4\pi}{(2\pi)^{3}}
\frac{1}{k^2-p^2}
\frac{1}{pk}
\sum_i \lambda_i
u_k(r_i)
\sin(pr_i)
\end{equation}
In the next section we present plots of the 
high-momentum radial wave function,
$\tilde\Phi_{\nk}(p)$,   
defined as
\begin{equation}
\langle \np|\psi_{l=0}\rangle = 
\frac{\tilde\Phi_{\nk}(p)}{k}Y_{00}(\hp)
\end{equation}
while the high-momentum pair density will be proportional to the 
square $|\tilde\Phi_{\nk}(p)|^2$.

\begin{figure*}
\centering
\includegraphics[bb = 50 420 550 780]{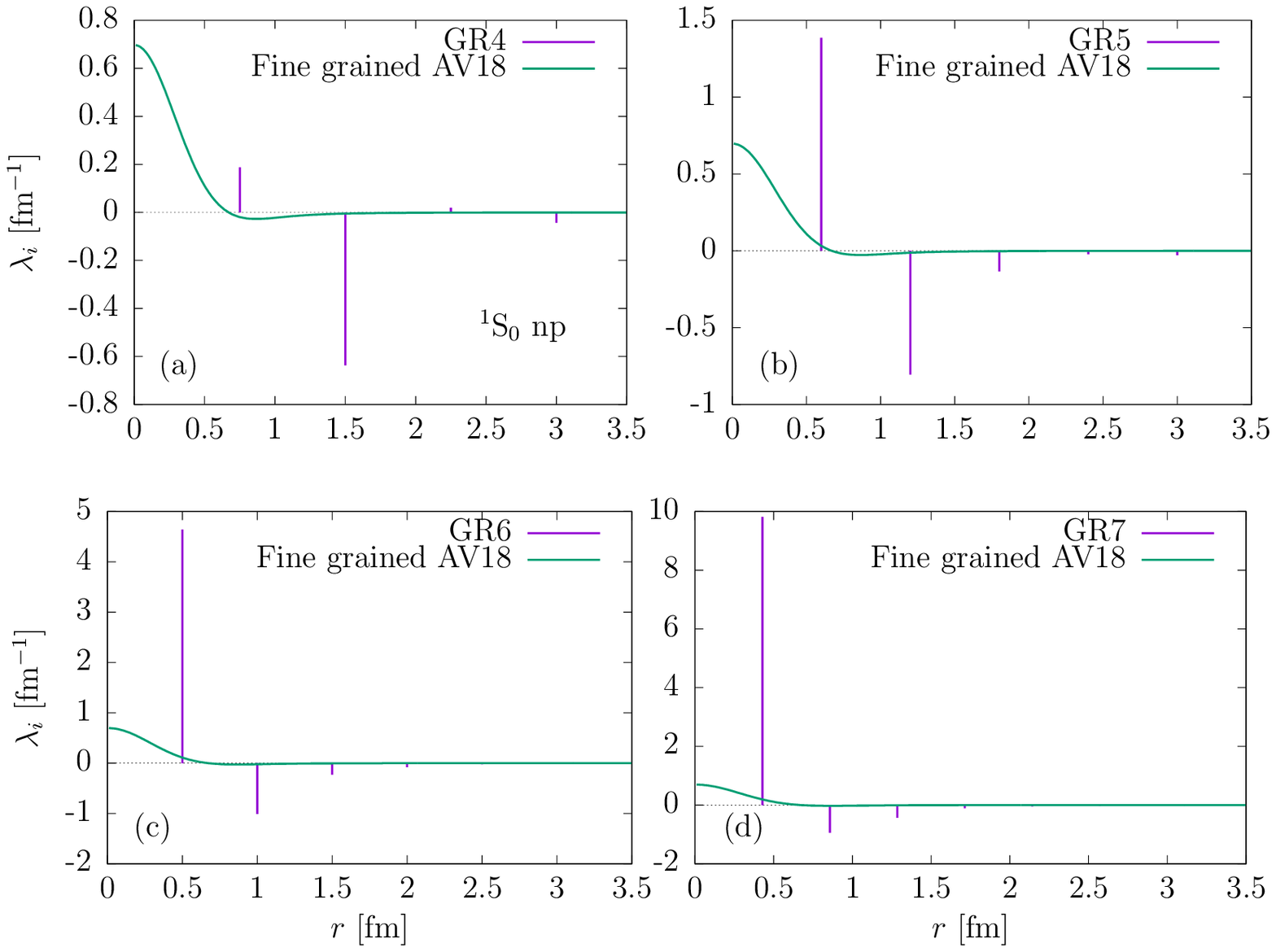}
\caption{Comparison between the fine grained AV18
potential \cite{Wiringa:1994wb} for the ${}^{1}$S${}_{0}$
partial wave with 600 delta shells and the corresponding coarse graining 
 with 4, 5, 6 and 7
(from left to right and top to bottom, respectively)
equally spaced delta shells between 0 and 3 fm.}\label{figpot}
\end{figure*}

\subsection{G-matrix}

The G-matrix is defined by
\begin{equation}
V|\psi_{\nk}\rangle= 
G|\nk\rangle.
\end{equation}
  From Eq. (\ref{matrix-element}) it is straightforward to obtain 
the $G$-matrix element for a back-to-back nucleon 
pair in the $^1S_0$ channel,
 as  
\begin{eqnarray} \label{G-matrix}
G(p,k)= 
\langle \np | G_{^1S_0} | \nk\rangle =
\frac{\sum_{i=1}^N \lambda_i   u_k (r_i) \sin(p r_i)}{2\pi^2 M_N p k  } 
\end{eqnarray}
Note that this expression has been obtained for $p>k_F$, and then
analytically extended to the $p< k_F$ branch.  This matrix element
quantifies the short-range correlations in the initial back-to-back
nucleon pair state with relative momentum $k$, allowing a transition
to a state with momentum $p$. As it is known, the diagonal elements
$G(k,k)$ for $k \le k_F$ contribute to the nuclear binding energy,
whereas the off-diagonal elements with $k \le k_F$ and $p> k_F$
correspond to induced high momentum components above the Fermi level
produced  by the NN interaction.

\begin{figure*}
\centering
\includegraphics[bb = 70 535 596 770]{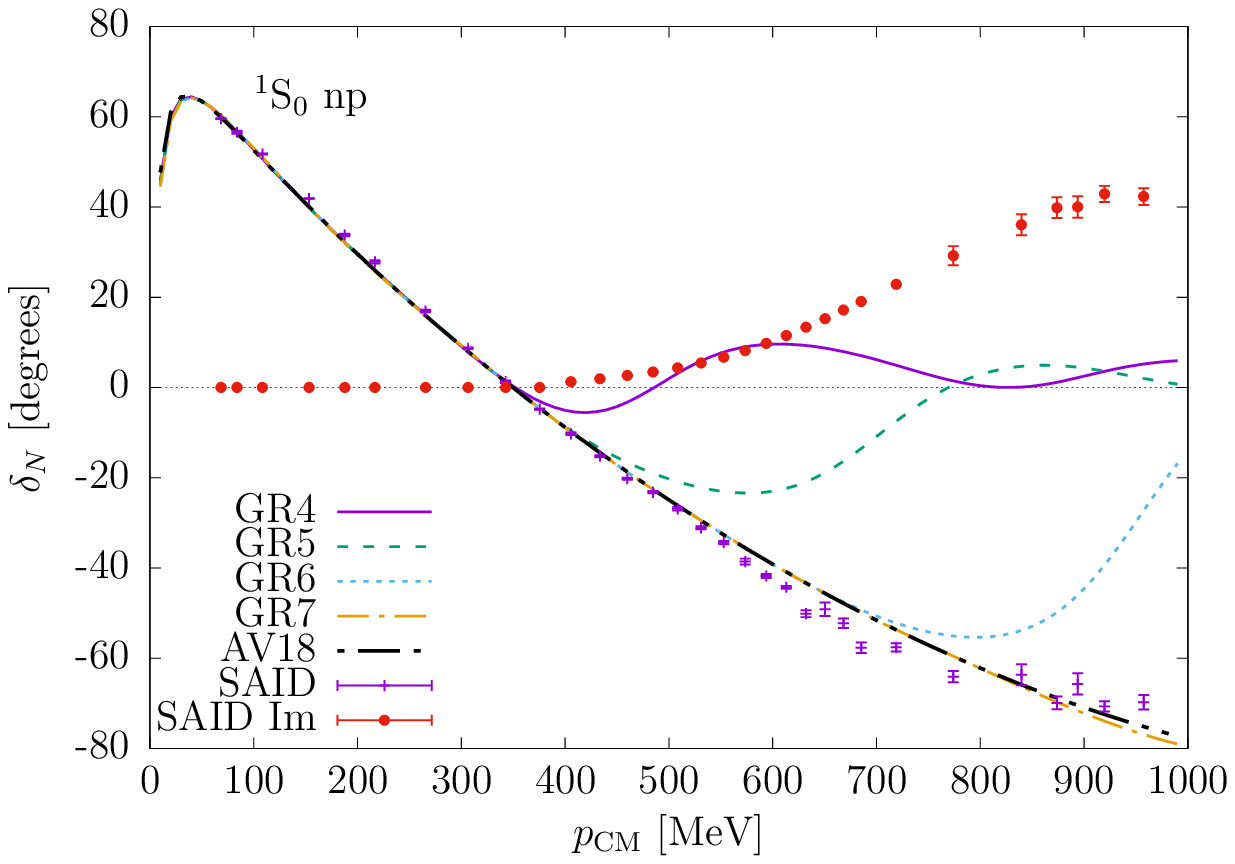}
\caption{Neutron-proton real phase-shifts in terms of the relative momentum 
in the CM system, 
$p_{\rm CM}=\sqrt{M_NT_{\rm LAB}/2}$,  for the ${}^{1}$S$_{0}$  partial
wave and for different number of delta shells, compared to
those obtained with the AV18 potential \cite{Wiringa:1994wb}.
We also plot the imaginary part of the phase shift. The data are  from the SAID 
database \cite{Arndt:2003fj}.}
\label{phaseshifts}
\end{figure*}

\begin{table*}[htp]
\begin{center}
\begin{tabular}{|c|c|c|c|c|}
\hline
$\lambda_i$ (fm${}^{-1}$) & 4 DS ($\Delta r=0.75$ fm) & 
5 DS ($\Delta r=0.6$ fm) & 6 DS ($\Delta r=0.5$ fm) & 
7 DS ($\Delta r=0.43$ fm)\\
\hline
$\lambda_1$ & 0.19 & 1.39 & 4.64 & 9.82\\
\hline
$\lambda_2$ & -0.636 & -0.81 & -1.011 & -0.937\\
\hline
$\lambda_3$ & 0.019 & -0.13 & -0.231 & -0.433\\
\hline
$\lambda_4$ & -0.044 & -0.023 & -0.081 & -0.107\\
\hline
$\lambda_5$ & - & -0.028 & -0.022 & -0.052\\
\hline
$\lambda_6$ & - & - & -0.018 & -0.018\\
\hline
$\lambda_7$ & - & - & - & -0.016\\
\hline
\end{tabular} \caption{Strengths 
$\lambda_i$ of the delta-shell (DS) potential (in fm${}^{-1}$) fitted
  to reproduce the same phase shifts as the AV18 potential
  \cite{Wiringa:1994wb} for the $^1S_0$ partial wave.  The number of
  DS in the range $0-3$ fm increases from left to right.  In each
  column the separation ($\Delta r$) between consecutive delta shells
  is also displayed.}\label{strengths}
\end{center}
\end{table*}

\section{Numerical Results} \label{sec3}

We come to our numerical results. In this initial exploratory study we
are interested in the properties of the solution of the B-G equation
for several NN potentials in $S$-wave.  In particular we compare
several coarse-grained potentials with the results obtained with the
AV18 interaction \cite{Wiringa:1994wb}.  In a previous study
\cite{NavarroPerez:2011fm} we performed a coarse graining of the AV18
potential for $p_{\rm max} \sim 2 \, {\rm fm}^{-1}$ with $N=5$ delta
shells separated by $\Delta r \sim 0.5 {\rm fm}$.  Here we extend the
analysis to higher energies. We adjust the strength of the delta
shells below $r_{\max}=3$ fm in order to reproduce the phase-shifts of
the AV18 potential for the ${}^{1}$S$_0$ partial wave, up to some
maximum LAB kinetic energy $E_{\rm max}$.  By increasing $E_{\rm max}$
we find that the number of delta shells incorporated has to be
accordingly increased.

 In figure \ref{figpot} we show the fine grained version of the AV18
 \cite{Wiringa:1994wb} potential for the $^1$S$_0$ partial wave
 compared to the coarse grained potentials for the same partial wave with
 4, 5, 6 and 7 delta shells, respectively 
(these potentials which will be denoted as GR4, GR5, GR6 and
 GR7 respectively). The values of the strengths of the delta shells
 potentials can be read from table \ref{strengths}. In each case, the
 values of $\lambda_i$  have been adjusted to reproduce the same
 phase shifts as those of the AV18 potential up to a certain kinetic
 energy in LAB, which is increasing with the number of delta shells
 between 0 and 3 fm.
 The AV18 potential has been sampled at 600 equally
 spaced points between 0 and 6 fm ($\Delta r= 0.01$ fm),
 and then the equivalent strengths of the 600 delta shells 
have been calculated.
 
 It can be seen from figure \ref{figpot} that the strength of the
 first delta shell (second row) increases when the number of delta
 shells increases, or equivalently when the distance between
 consecutive shells decreases. This is necessary in order to reproduce
 the same phase shifts as those of the AV18 with a small number of
 delta shells. The strength of the first delta shell mimics the
 repulsive properties of the NN potential at short distances. When we
 increase the number of deltas, we reveal the short-distance behavior
 of the potential, and the relevance of the repulsive short-distance
 region becomes more and more important.

\begin{figure*}
\centering
\includegraphics[bb = 50 420 550 780]{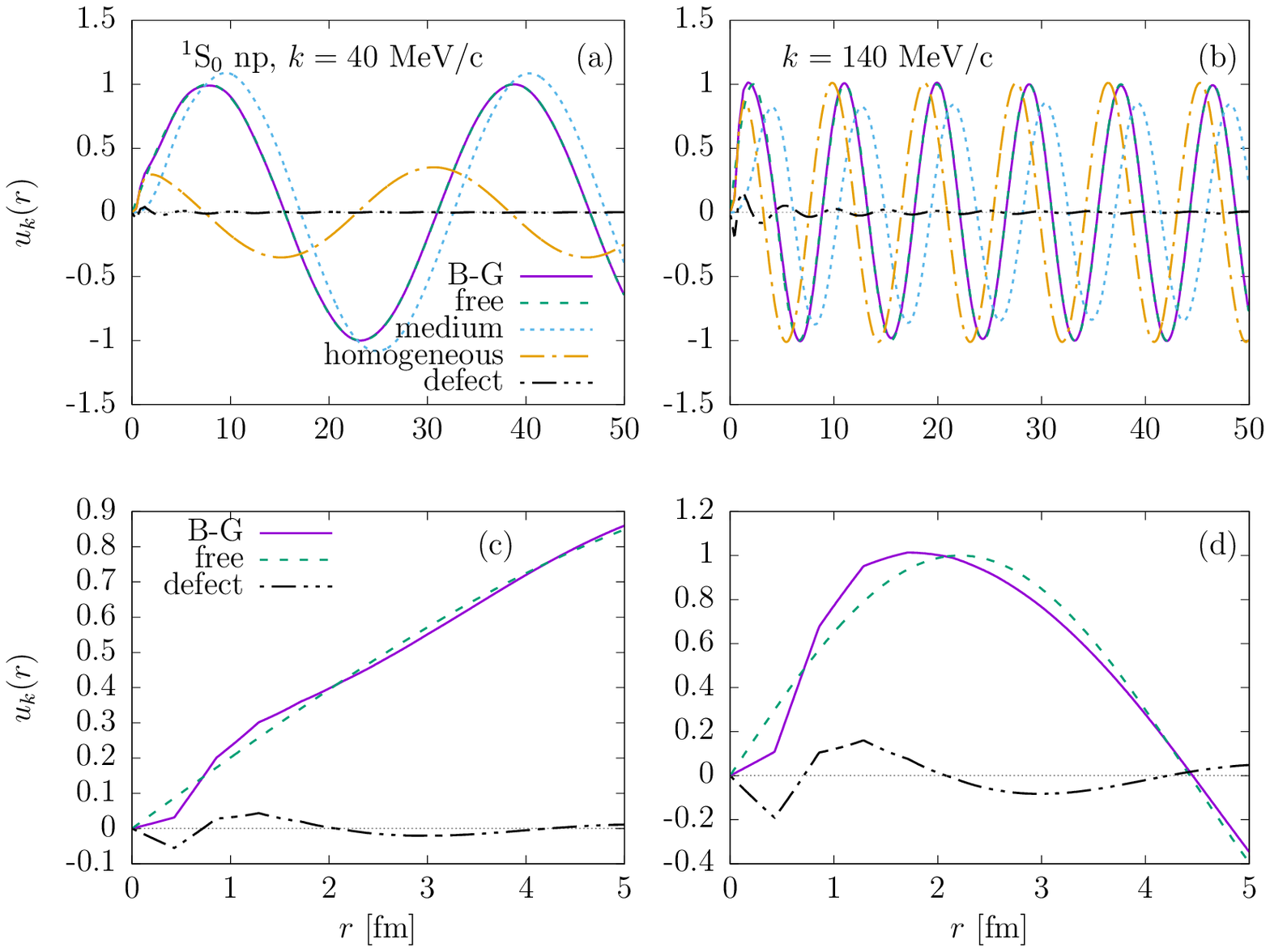}
\caption{Top panels: Comparison between the reduced wave functions
  $u_k(r)$ for two different ($k=40$ MeV/c on the left panel and
  $k=140$ MeV/c on the right one) initial relative momenta of the
  nucleon pair. We use the GR7 interaction corresponding to 7 DS potential
  with parameters given in the last column of table
  \ref{strengths}. The curves correspond to the Bethe-Goldstone wave
  function (\ref{whole_solution}), the free wave solution $\sin(kr)$,
  the medium correction or particular solution
  (\ref{particular_solution}), the homogeneous wave function given by
  eq. (\ref{homogeneous_sol}) and, finally, the defect wave function,
  which is defined as $\Delta u_k(r)= u_k(r)-\sin(kr)$.  Bottom
  panels: detail of the BG, free and defect functions in the region
  below 5 fm.  
Note that the BG and the free wave functions are almost 
identical except for short distances and they cannot be disentangled
in the upper panels. Their difference can  be appreciated 
in the bottom panels for short distances.
}\label{defectgraph}
\end{figure*}

All the potentials shown in figure \ref{figpot} give the same real
phase shifts up to a certain CM momentum or kinetic energy in LAB
system.  This can be observed in figure \ref{phaseshifts}, where the
real phase shifts for the $^1$S$_0$ neutron-proton (np) partial wave
are plotted for all the potentials considered in this work.  The delta
shell strengths have been adjusted to reproduce the same phase shifts
as the AV18 potential. The adjusted region in LAB energy is increased
with the number of delta shells.  For instance, for a potential with 4
delta shells (DS), we have fitted the phase shifts up to $T_{\rm
  LAB}=250$ MeV; while for 6 DS, the fitted region ranges up to $T_{\rm
  LAB}=921$ MeV.  It can also be seen that a potential with 7 DS
between 0 and 3 fm reproduces the phase-shifts of the AV18 potential
up to almost $T_{\rm LAB}=2$ GeV.  For completeness we also show in
figure \ref{phaseshifts} the imaginary part of the phase-shifts which
differs substantially from zero above $\Delta$ production threshold,
$p_\Delta \sim 550$ MeV, whereas it is exactly zero in the case of the 
 AV18 and  GR potentials.

\begin{figure*}
\centering
\includegraphics[bb = 50 610 550 780]{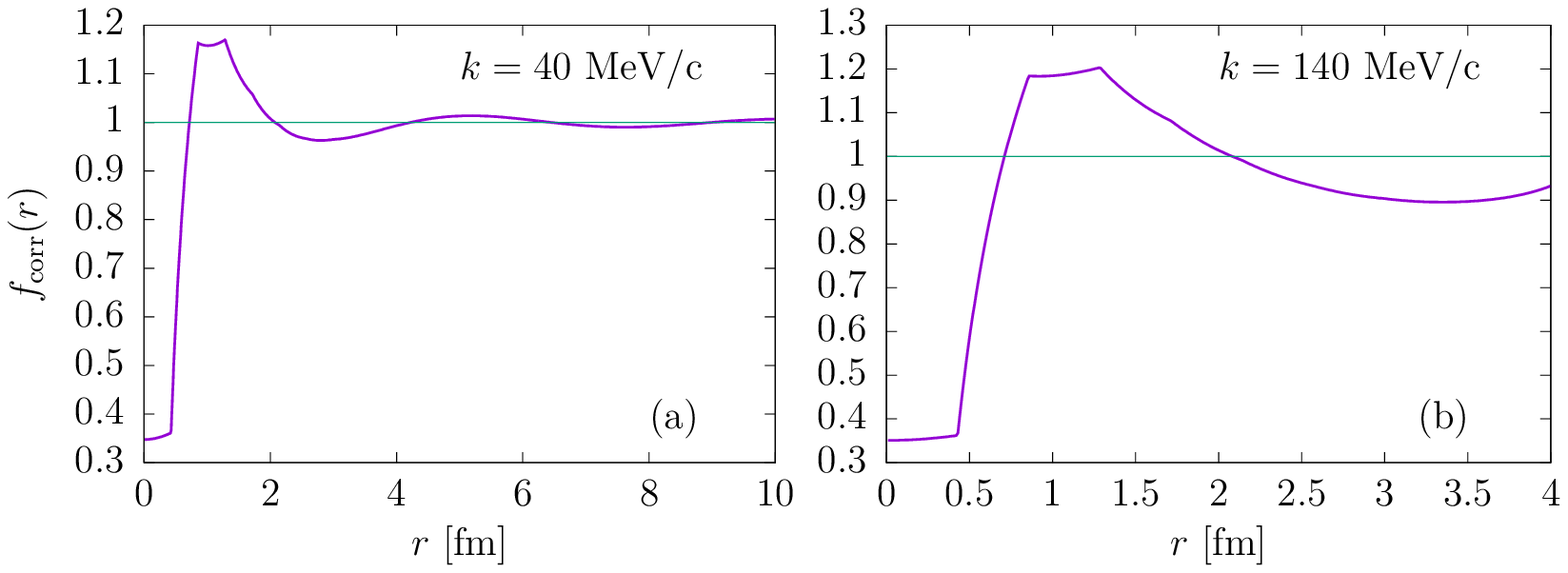}
\caption{Correlation function $f_{\rm corr}(r)$, defined as 
the quotient between the BG correlated wave function $u_{k}(r)$
and the free unperturbed wave function $\sin(kr)$, for two
different values of the relative momentum of the pair, $k=40$
and $k=140$ MeV/c, respectively. These results have been obtained
for the GR7 interaction with 7 DS.}\label{corrfunc}
\end{figure*}

In figure \ref{defectgraph} we plot the solution, $u_k(r)$, of the BG
equation for two values of the relative momentum of the nucleon pair.
These results have been obtained using the GR7 potential corresponding
to 7 DS. From the figure we observe that the BG wave function and the
free one are almost identical because the effects of the potential get
damped at large distances. The BG solution do not present any
phase-shift because the presence of the medium prevents the scattering
into already occupied states with $k<k_F$.  The effect of the
interaction is to distort the wave function at short distances,
producing a bending with rapid oscillations of small amplitude.  These
small, fast oscillations are a direct signal of the presence of
high-momentum components and the leading feature of short-range
correlations.  The bending of the wave function at short distances can
be better appreciated in the lower panels of fig. \ref{defectgraph}.
The rapid oscillations can be observed as ``spikes'' in the BG wave
function and in the defect wave function defined by
\begin{equation}
\Delta u_k(r)=u_k(r)-\sin(kr).
\end{equation}
 The spikes are due to the singular delta shell potential which
 introduces discontinuities in the derivative at the points $r_i$. One
 can also observe the oscillatory and amplitude-decreasing behavior of
 the defect wave function, making the BG solution to oscillate around
 the free solution with decaying amplitude for long distances.

 From our results we can also obtain the value of the healing
 distance, defined as the point where $\Delta u_k(r)$ first vanishes
 \cite{Fet71}.  It is almost independent on the relative momentum of
 the pair.  By inspection of the lower panels of figure
 \ref{defectgraph} it is $\sim 0.75$ fm. This value for the healing
 distance is smaller than the one of a hard-core potential at $r_c=
 0.4$ fm, which is $\sim 1.34$ fm \cite{Wal95,Fet71}. This makes sense
 because in a hard-core potential at $r_c=0.4$ fm the wave function is
 forced to be zero at $r_c$ and not at $r_0=0$ as here.  Therefore,
 one would expect the necessary distance to \textit{heal} to the
 unperturbed wave function to be larger for the hard-core potential
 case.

In the upper panels of figure \ref{defectgraph} we also show the
solution $u_h(r)$ for the homogeneous ODE (\ref{homogeneous_ODE}), and
the medium effect or particular solution $u_p(r)$ of
eq. (\ref{particular_solution}). It can be seen that both solutions
are shifted with respect to the free wave function and that they have
different amplitudes, but in such a way that their sum has no phase
shift with respect to the free solution at large distances.

In figure \ref{corrfunc} we show the correlation function, defined as
the quotient between the correlated and uncorrelated wave functions
\begin{equation}
f_{\rm corr}(r)=\frac{u_{k}(r)}{\sin(kr)},
\end{equation}
for two different relative momenta of the nucleon pair, $k=40$ and
$k=140$ MeV/c, respectively. We only show the short-distance region
region before the first zero of the uncorrelated wave function
$\sin(kr)$, reached for $r=\pi/k$.  While there is no phase-shift
between both wave functions, this is the case for very large distances
only.  In figure \ref{corrfunc} it can be seen that the deviations
from unity of the correlation function primarily occur at very short
distances, for $r< 0.5$ fm, where it takes small values. For larger
values, around 1 fm, the correlated wave function \textit{heals} and
shows a trend to oscillate around the unity with damping
amplitude. Thus the correlation function approaches oscillatory to
unity.

\begin{figure*}
\centering
\includegraphics[width=12cm, bb = 120 280 520 800]{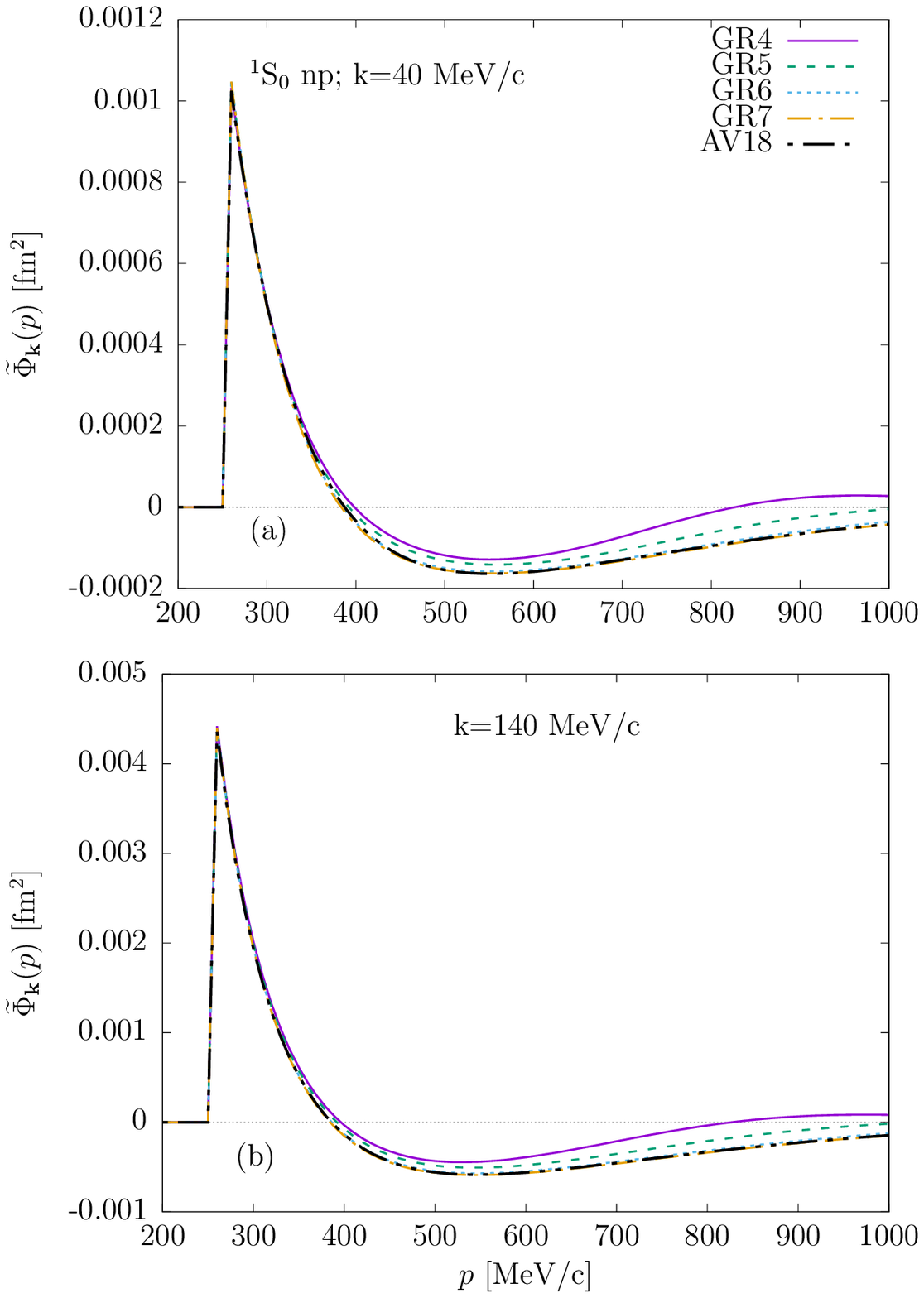}
\caption{High-momentum radial wave function $^1S_0$ in momentum
  representation for two values of the relative momentum $k$ of the
  back-to-back correlated pair.  The results for different coarse
  grained potentials are compared to the ones of the AV18 potential.}
\label{momdist}
\end{figure*}

In figure \ref{momdist} we show the high-momentum radial $^1S_0$ wave
function $\widetilde{\Phi}_{\nk}(p)$ of the back-to-back correlated
pair, for relative momenta $k=40$ and $k=140$ MeV/c, respectively,
computed with several coarse grained potentials, and compared to the
AV18 potential.  The square of this function will contribute to the
high-momentum tail induced by the short-range correlations in the
momentum distribution.

 From the figure we observe that, above the Fermi momentum, which has
 been fixed in this work to $k_F=250$ MeV/c, and below p = 400 MeV/c,
 all the potentials basically give the same high-momentum
 tail. Moreover the results between $p=400$ and 600 MeV are very
 similar, and above $p=600$ their differences start to be more
 pronounced. Going back to figure \ref{phaseshifts}, we can see
 that a common feature of these potentials is that they produce the
 same phase-shifts for LAB energy below 250 MeV.  From these results,
 we conclude that the phase-shift information above 250 MeV in kinetic
 LAB energy is irrelevant to properly describe the high-momentum tail
 up to 400 MeV/c.

\begin{figure*}
\centering \includegraphics[bb = 120 280 520 800]{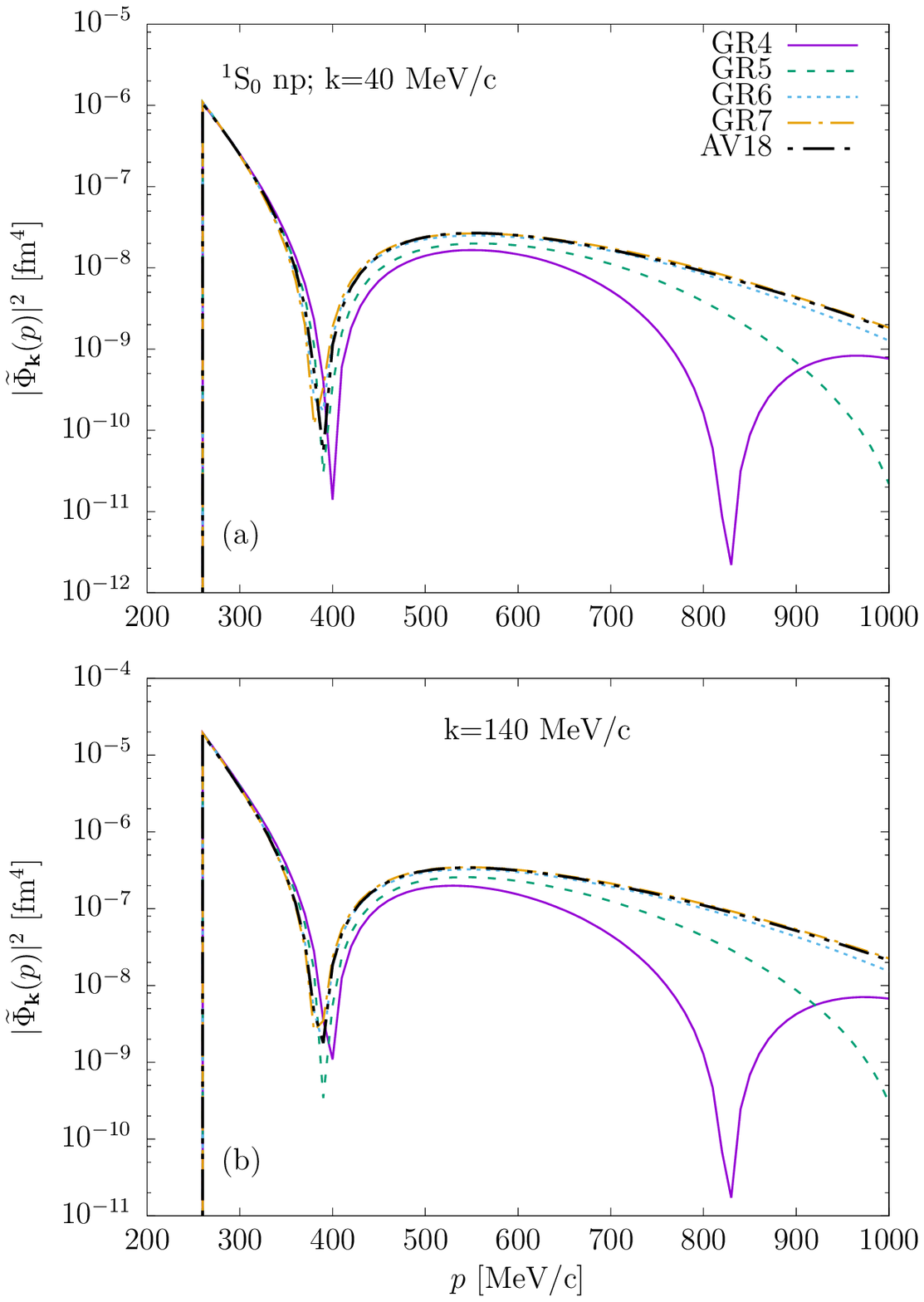}
\caption{The same as figure \ref{momdist}
for the square of the high-momentum tail of the 
radial wave function in momentum representation.
}\label{momdistsquared}
\end{figure*}

A similar case arises when comparing the results of the potentials GR6,
GR7 and AV18.  These three interactions give essentially the same
high-momentum tail up to 1000 MeV/c.  When looking at figure
\ref{phaseshifts}, one can observe that these three potentials
describe the same phase shifts up to 1000 MeV in LAB energy, thus
suggesting again that additional information contained in the
phase-shifts beyond that energy has no influence in the high-momentum
tail of the correlated wave function below $p= 1000$ MeV/c.

A closer view of the differences between the high momentum tail
induced by the different interactions is provided in figure
\ref{momdistsquared}, where we show the square of the momentum wave
function $\left|\widetilde{\Phi}_{\nk}(p)\right|^2$, plotted in
logarithmic scale.  This function is proportional to the high momentum
distribution of the back-to-back nucleon pair in relative
${}^{1}$S${}_{0}$ state, where the well known minimum at $\sim 400$
MeV/c, produced by the node of the wave function, is apparent.

\begin{figure*}
\centering
\includegraphics[bb = 70 535 596 770]{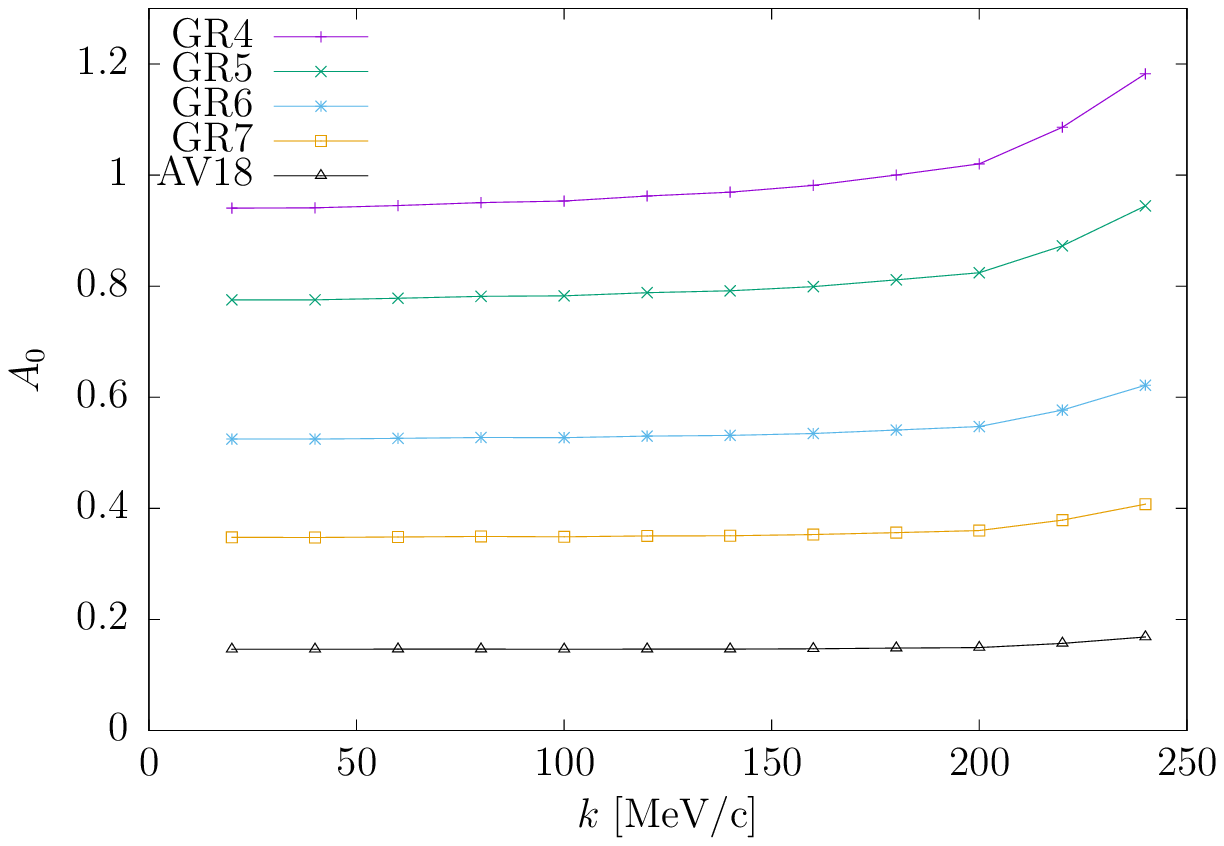}
\caption{The values of the correlation function at the origin 
$A_0= f_{\rm corr}(0)$
for the different potentials considered in this work.
}\label{renorm}
\end{figure*}

An important quantity for the assessment of the short range
correlations is the value of the correlation function at the origin,
provided in our formalism by the constant $A_0$
\begin{equation}
A_0= f_{\rm corr}(0) \equiv \lim_{r\rightarrow0} \frac{u_k(r)}{kr}
\end{equation}
In figure \ref{renorm} we show the values of $A_0$ for the different
potentials considered in this work. These values are plotted against
the initial relative momentum $k$ of the back-to-back nucleon pair.
In our approach the value of $A_0$ is computed in each iteration by
imposing the condition that the correlated wave function $u_k(r)$ goes to the
uncorrelated one $\sin(kr)$ for large values of $r$. 
The  values plotted in fig. \ref{renorm} correspond to the 
6th iteration to solve the BG equation, after convergence is reached.  
The importance of these curves lies in the fact that they 
correspond to the values that should be given as the initial guess in
the iterative procedure to reach convergence in only one iteration.

By inspection of figure \ref{renorm} we see that for all the
potentials considered here, the $A_0$-values are quite stable in the
whole range of relative momenta $k$ up to the Fermi momentum,
$k_F=250$ MeV/c.  There is only a small enhancement at the end of the
curves when one is approaching the Fermi momentum. 

 The value $A_0$ measures the hardness of the potential at short
 distances. The larger the value of $\lambda_1$ (the amplitude of the
 first repulsive delta shell) the smaller the value of $A_0$, 
 and  the  smaller the wave function at short distances.

\begin{figure}
\centering
\includegraphics[bb = 200 240 450 800]{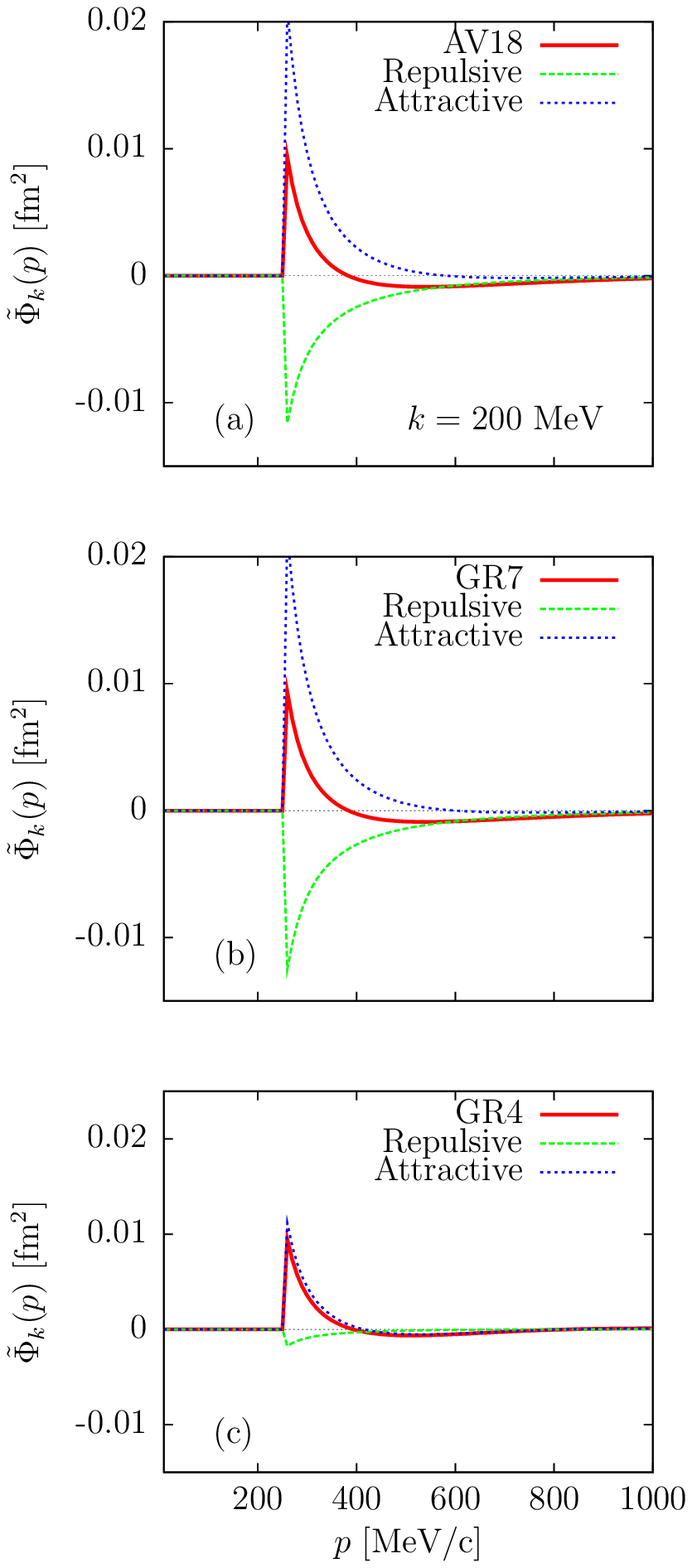}
\caption{Separation of the attractive and repulsive
high momentum components of the wave function for $k=200$ MeV/c 
for different  NN potentials.
}\label{rep-attrac}
\end{figure}

It is instructive to separate in Fig.~\ref{rep-attrac} the
contribution of the attractive and repulsive parts of the potential to
the high momentum tail of the wave function. For the coarse grained
potentials, the first delta shell is positive and repulsive, while the
others are negative and attractive. According to 
Eq. (\ref{momentum-wave}) the potential enters into the momentum 
tail through the sum
$\sum_i \lambda_i u_k(r_i) \sin(pr_i)$. 
We thus can separate this sum into two terms including only the repulsive 
$\lambda_i > 0 $ and attractive $\lambda_i < 0$ parts  of the potential.

The results of this separation are shown in Fig.~\ref{rep-attrac} 
for  $k=200$ MeV. As can be seen in the case of the AV18 potential, where
$V(r)> 0$ for $r< 0.75 {\rm fm}$ and $V(r)<0$ otherwise, the repulsive
short-distance piece including the core gives a positive and large
momentum tail, which is partly compensated by the negative contribution
stemming from the attractive longer range potential.  This effect is
reproduced with the coarse grained GR7 potential which is equivalent,
as shown before, to the AV18 potential both for the phase-shift and
for the high-momentum wave function.  

However, this feature depends strongly on the resolution scale $\Delta
r$. In the bottom panel of the figure we observe that, in the case of
the GR4 potential, the repulsive contribution is much smaller than the
attractive contribution which gives alone almost the same total
momentum tail as the AV18 or GR7 potentials. In fact most of the
high-momentum tail is here dominated by the second, attractive delta
located at $r=1.5$ fm.  This remarkable feature suggests that the
short-range correlations can be traced back primarily {\em not} to the
traditional repulsive core phenomenology but rather to the
{\em attractive mid-range} part of the interaction. This opens a gateway to
a perturbative treatment of short-range correlations which will be
exploited elsewhere.

Our analysis also explains a feature which has been systematically
found in large scale calculations, namely the appearance of an
universal diffraction minimum at $p=400$ MeV in the $^1S_0$ momentum
distribution \cite{Wiringa:2013ala}.  Although this is known to be
produced by the mid-range part of the interaction and not to the
repulsive core, we believe its diffractive origin can be understood in
simple terms; our analysis shows that it is due to the fact that in
the GR4 potential the high-momentum tail in the wave function is
dominated by the second delta shell, located at $r_2=1.5 {\rm
  fm}$. According to Eq.(\ref{momentum-wave}) the contribution in the
S-wave is proportional to $\sin(pr_2) $ which vanishes for $p= \hbar c
\pi/r_2 \approx 400 {\rm MeV}$.

Let us mention that the momentum representation given in
Eq. (\ref{momentum-wave}), suggests a simple way of parameterizing the
high momentum components of the wave function.  The short range
correlations are encoded into the quantities $c_i(k) \equiv \lambda_i u_k
(r_i)$, $i=1,\ldots,N$, which in principle could be 
parameterized with smooth functions which do not depend strongly on $k$.
 This could open an interesting line of research to
investigate if information about these 
``coarse-graining correlation functions'', $c_i(k)$, could be
experimentally extracted for example in two-nucleon knock-out
experiments such as $(e,e'NN)$.

\begin{figure}
\centering
\includegraphics[width=8cm, bb=190 600 430 780]{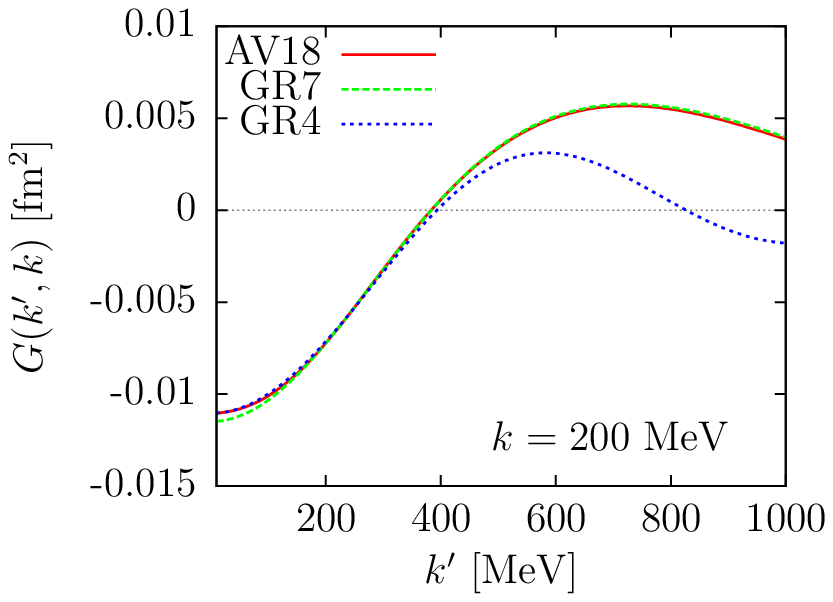}
\caption{ G-matrix in the $^1S_0$ channel for a back-to-back pair with
  initial relative momentum $k=200$ MeV/c.
 We compare the results
  corresponding to the three potentials AV18, GR4 and GR7.
}\label{fig-G-matrix}
\end{figure}

\begin{figure*}
\centering
\includegraphics[width=14cm , bb=80 240 550 790]{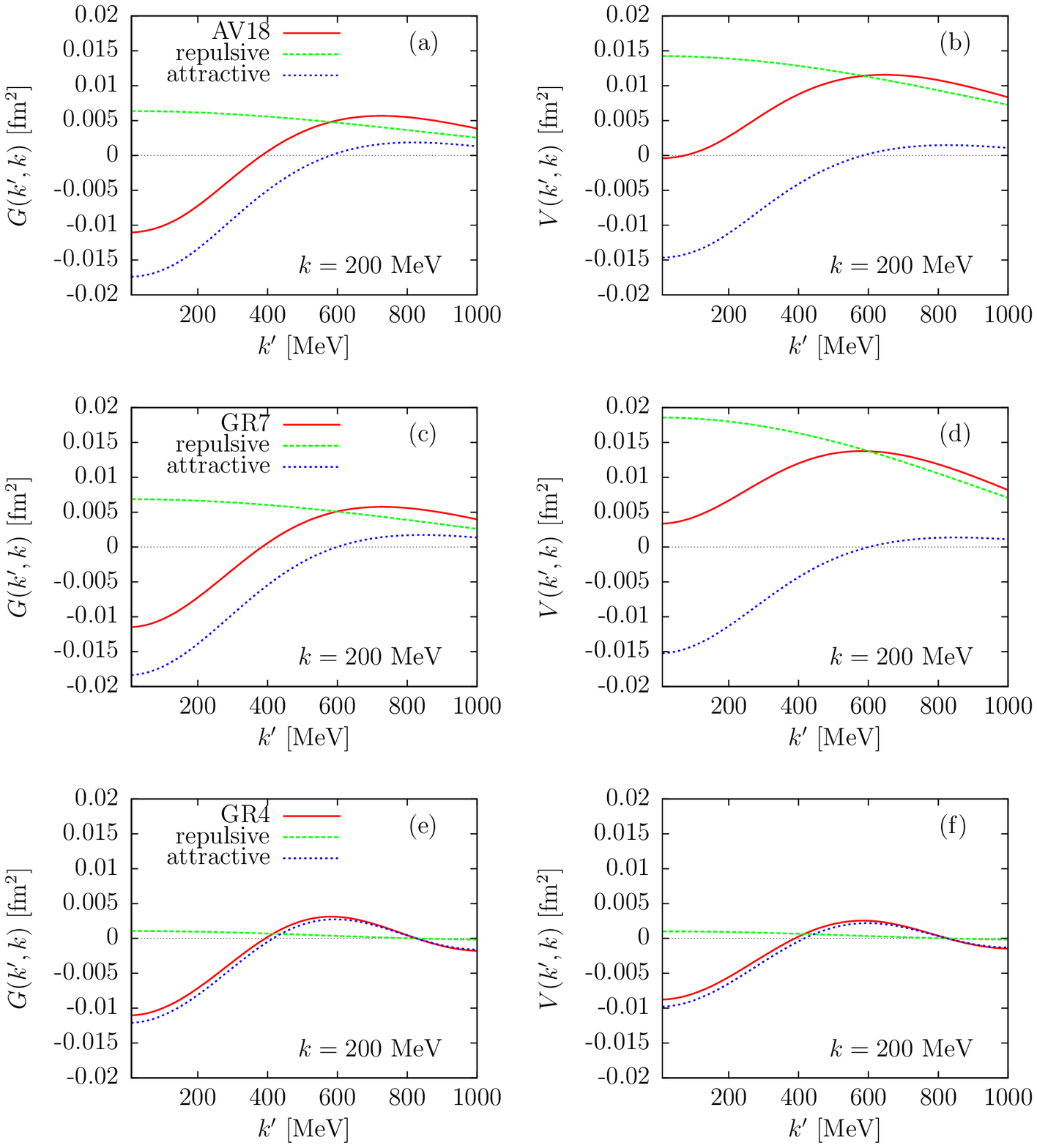}
\caption{
G-matrix in the $^1S_0$ channel 
for a back-to-back pair  with initial relative momentum 
$k=200$ MeV/c (left panels). 
 We compare the results
  corresponding to the three potentials AV18, GR4 and GR7.
In the right panels we compare corresponding potential matrix element 
in momentum space. In al cases we also show the separate contribution
from the repulsive and attractive parts of the potential. 
}\label{G-matrix2}
\end{figure*}

\begin{figure}
\centering
\includegraphics[width=8cm , bb=200 420 420 780]{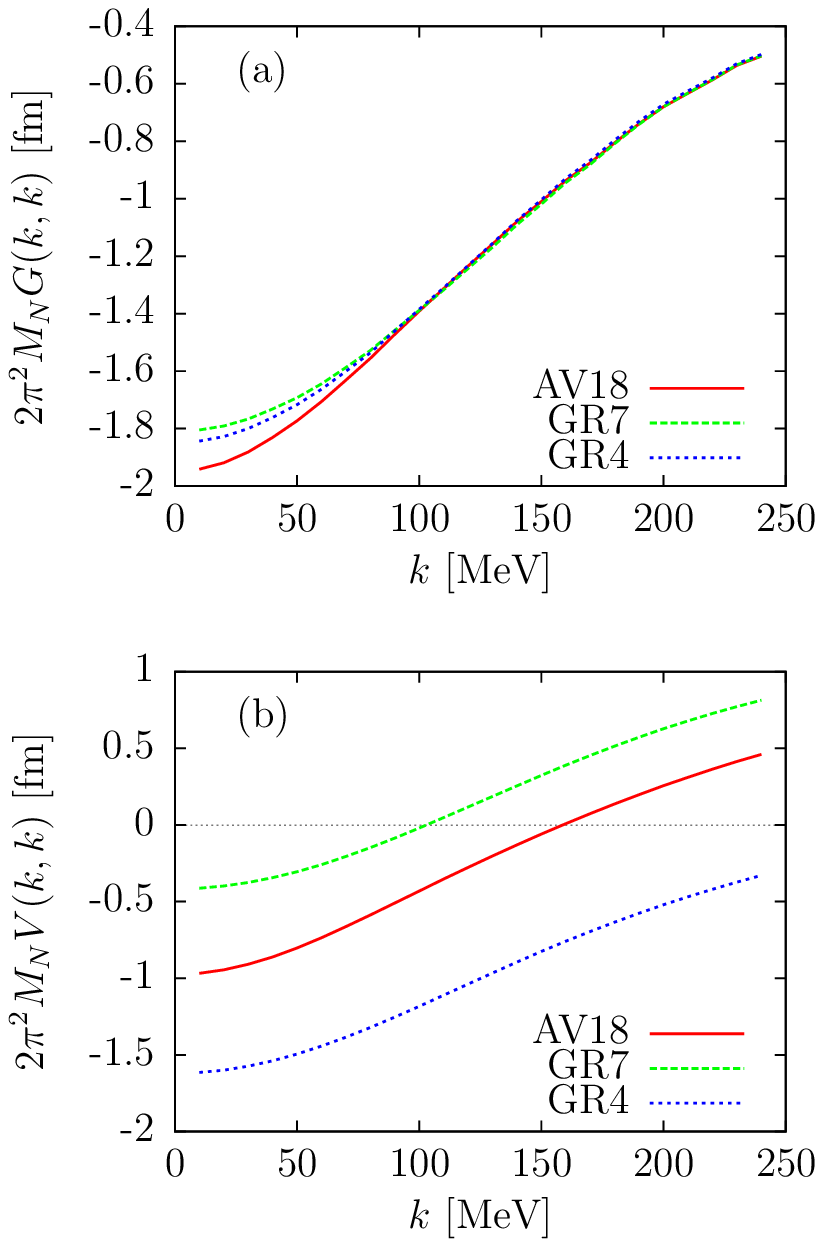}
\caption{
Upper panel: 
diagonal G-matrix (multiplied by $2\pi^2 M_N$) in the $^1S_0$ channel 
for a back-to-back pair  with initial relative momentum 
$k=200$ MeV/c. 
 We compare the results
  corresponding to the three potentials AV18, GR4 and GR7.
Bottom panel: the same for the  potential matrix element 
in momentum space.
}\label{G-diagonal}
\end{figure}

\begin{figure}
\centering
\includegraphics[width=8cm , bb=200 610 420 780]{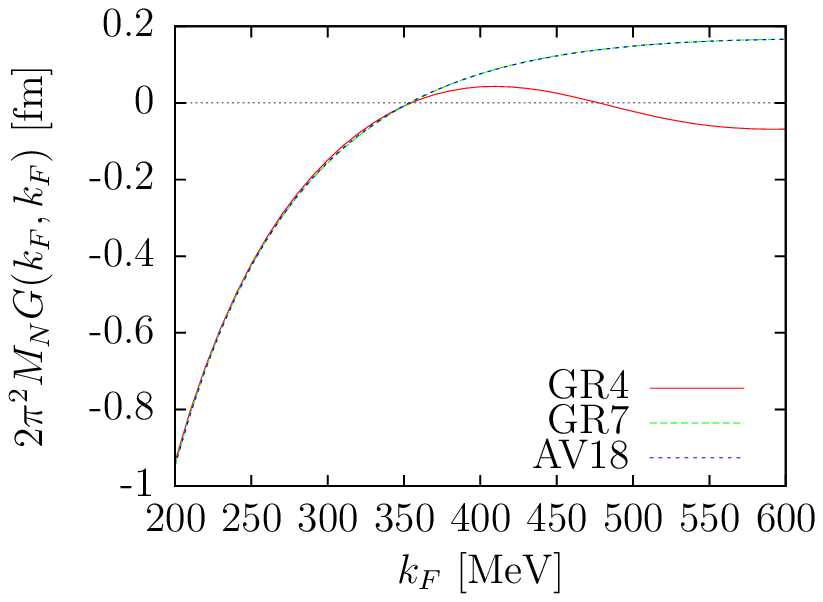}
\caption{ Diagonal G-matrix (multiplied by $2\pi^2 M_N$) in the
  $^1S_0$ channel for a back-to-back pair with $k=k_F$ as a function
  of $k_F$.  We compare the results corresponding to the three
  potentials AV18, GR4 and GR7.
}\label{GkF}
\end{figure}

The coarse-grained approach allows us to compute easily the $G$-matrix
in the $^1S_0$ channel using Eq. (\ref{G-matrix}).  In
Fig.~\ref{fig-G-matrix} we show the momentum space representation
$G(k',k)$ for fixed momentum $k=200$ MeV/c, as a function of $k'$. As
we see the AV18 and GR7 potentials give essentially the same
$G$-matrix up to $k'=1000$ MeV/c. The GR4 potential agrees well with
both of them for $k' \le 500 \,\rm MeV = 2 k_F $, and start to
considerably differ for $k' \sim 600$ MeV/c.

It is a striking result that, being the three potentials (AV18, GR7
and GR4) so different in momentum space, they generate essentially the
same G-matrix for momenta below $2k_F$.  This is because the
$G$-matrix embodies the short-range correlations through the already
mentioned coarse-grained correlation functions $c_i(k)=\lambda_i
u_k(r_i)$.  As we have seen the information encoded in these functions
can be separated into repulsive and attractive contributions that, 
in  the average, are very similar for all the potentials
considered here. 

A deeper insight is provided in fig. \ref{G-matrix2}, where we  
show the separate repulsive and attractive parts of the 
NN potential in momentum space 
$V(k',k)$, and its corresponding contributions to the $G$-matrix. In the case of the AV18 potential the repulsive and attractive G-matrices 
are large and opposite in sign and partially cancel in the total $G$-matrix.
The same can be said in the case of the GR7 potential, that gives essentially the same repulsive and attractive $G$-matrices as the AV18 potential.

However the GR4 potential provides a different case, where the
repulsive contribution to the $G$-matrix is negligible, while the
attractive part of the potential generates alone almost the full
$G$-matrix.  Note also that in the case of the AV18 and GR7
interactions, $G(k',k)$ is very different from $V(k',k)$. However the
GR4 potential is very similar to its $G$-matrix. In fact, this
suggest once more that one could use directly the GR4 interaction to
investigate the role of short-range correlations in perturbation
theory.

For momenta below $\sim k_F$, one expect that the diagonal $G$-matrix,
$G(k,k)$, to be identified with the effective NN interaction inside
the nucleus, be also quite stable for all the interactions.  This is
shown in Fig. \ref{G-diagonal}, where we compare $G(k,k)$ and the NN
potential $V(k,k)$.  for the three interactions AV18, GR7 and GR4 in
the $^1S_0$ channel.  We have multiplied these functions by a
normalization factor so that these curves can be compared with the
usual definition of the $V_{\rm low k}$ potential
\cite{Bog03,NavarroPerez:2011fm}, which has already been  proposed as
a convenient effective interaction in the nucleus, and that in fact is
very similar to our results in figure \ref{G-diagonal}.

Finally, in figure \ref{GkF} we show the diagonal $G$-matrix computed
for $k=k_F$ as a function of the Fermi momentum, in a range that covers
the Fermi momentum of finite nuclei, nuclear matter, and extends well
above the $k_F$ values expected for neutron stars. The $G$-matrix
close the Fermi surface is related to the effective interaction of the
Landau-Migdal finite Fermi systems, and it should describe the
particle-hole interaction near the Fermi surface.  The GR4
potential  give the same results as the ones of 
GR7 and AV18 for $k_F < 350$ MeV, while
they start to disagree for larger $k_F$-values.

\section{Conclusions} \label{sec4}

In this work we have explored the role of coarse grained interactions
in the study of short-range correlations in nuclei, in terms of the
high momentum components in the back-to-back nucleon pair wave
functions.  We remind that these components have traditionally been
linked to the existence of a strong repulsive core between nucleons
located at about $r_c \sim 0.3-0.4 {\rm fm}$, preventing any kind of
perturbative treatment, and promoting large scale calculations to
befit the complexity of the problem. On the other hand, for a maximum
back-to-back momentum $p$ there corresponds a limiting resolution
wavelength $\Delta r $ which effectively samples the nuclear wave
function as well as the corresponding NN interaction above the
repulsive core scale $r_c$. Therefore, we expect some cancellation
between the repulsive and attractive parts of the interaction in the
region around the core for $ p r_c \lesssim 1$, which means $p
\lesssim 600 {\rm MeV}$. Previous experience fitting and selecting NN
data in high precision partial wave analysis suggests that this
``moderately'' high momentum and core-blind regime might
exist. Conversely, this also seems to imply that genuine
repulsive-core features start being visible for $p \gtrsim 600 {\rm
  MeV}$.

Therefore, as a first step in this exploratory work, we have addressed
the problem of solving the Bethe-Goldstone equation in the $^1$S$_0$
partial wave for several delta-shell potentials, including the
fine-grained version of the AV18 potential. The strengths of the
delta-shell potentials have been fitted to reproduce the same phase
shifts as the AV18 up to a certain kinetic energy in the LAB frame.

We have first developed a very efficient method to solve the
Bethe-Goldstone equation in coordinate space for this $^1$S$_0$
channel, obtaining convergence for the reduced wave function $u_k(r)$
after six iterations. With this solution we have studied the role of
short-range correlations to induce high momentum components in the
two-nucleon wave function.  We have also obtained two-nucleon momentum
distributions for this $S$-wave component and we have analyzed the
values $A_0$ of the radial wave function at the origin.  We have
compared our coarse-grained results with those obtained with AV18
potential, which embodies many of the traditionally assumed features
of the NN interaction, such as the repulsive core, and provides a good
description of NN phase-shifts up to np LAB energy of 2 GeV.

We find a few appealing features of the present approach.

\begin{itemize}

\item A huge reduction in the number of mesh points needed to solve
  the Bethe-Goldstone equation in coordinate space.  In the case of
  the AV18 we need to sample the interaction with 600 deltas (fine
  grain) to solve it, while in the coarse-grain case we just need a
  few mesh points (4--7) with almost the same result .  The
  coarse-grain interactions embodies all needed information on high
  momentum components of the nuclear wave function up to $p=600$--1000
  MeV, while accurately reproducing the AV18 $^1S_0$ phase-shifts up
  to LAB energies of 2 GeV. We plan to investigate this feature in
  more involved problem of the 
peripheral partial waves, including the effects of the tensor
  interaction in the $S=1$ channel.

\item We also witness a marginal contribution of the repulsive core so
  that, quite unexpectedly, a perturbative treatment of short-range
  correlations as observed in terms of high momentum components is
  envisaged. This is due to a suitable choice of the coarse graining
  scale which allows a dominance of the attractive contribution due
  to a reduction of the repulsive one.

\item We provide a simple explanation of the universally observed
  diffraction minimum in the high momentum distribution in the $^1S_0$
channel.

\end{itemize}

A pertinent comment triggered by figure \ref{phaseshifts} is that 
the input is dictated by the phase-shifts, but it is not clear 
up to which maximum energy they should be described by the potential 
to solve the Bethe-Goldstone equation without ambiguities.
The reason is because above 500--600 MeV in relative momentum, 
inelasticities due to production of nucleon resonances start
to contribute, producing a complex phase shift 
which cannot be described with a real potential.

Future plans include to solve the problem for higher
partial waves, including coupled-channels induced by the
nucleon-nucleon potential. This formalism could also 
be useful in the construction of G-matrices once 
the higher partial waves have been addressed.

\section{Acknowledgements} 
We thank A. Rios for useful communications.
This work has been partially supported by  Spanish Ministerio de
Economia y Competitividad and ERDF (European
Regional Development Fund) under contract FIS2014-
59386-P, by Junta de Andalucia grant FQM-225.
One of us, IRS, acknowledges support from a Juan de la Cierva
fellowship from MINECO (Spain).

\end{document}